\documentclass[12pt]{iopart}
\usepackage{iopams}
\usepackage{graphicx}
\expandafter\let\csname equation*\endcsname\relax
\expandafter\let\csname endequation*\endcsname\relax
\usepackage{amsmath}
\usepackage{amssymb}
\usepackage{cite}

\usepackage{xcolor}

\usepackage{dsfont}
\usepackage{bm}
\usepackage{verbatim}
\usepackage{enumerate}
\usepackage{url}

\newtheorem{remark}{Remark}

\begin{document}

\title[Far Tails of the Positional Distribution]{Far Tails of the Positional Distribution in random walk: A Saddle-Point Approach}
\author{Chen Zhou, YuDa Cheng, Yangji Li, Jing Xu and Wanli Wang}
\address{School of Mathematical Sciences, Zhejiang University of Technology, Hangzhou 310023, China}

\begin{abstract}
Rare events and large-deviation theory have attracted considerable interest in the study of random walks. Here, we investigate the rare events of the positional distribution of continuous-time random walks under ordinary and equilibrium initial conditions using the saddle point method. We show that the far-tail behavior is sensitive to the expansion of the waiting-time distribution near its minimum value. For a general power-series expansion near the minimum waiting time, the positional distribution exhibits an approximately exponential decay in its far tails. In contrast, when the waiting-time distribution has a cutoff at its minimum value, the positional distribution exhibits an unexpected Gaussian decay, in contrast to the far-tail behavior of the renewal-count distribution. We further find that the most likely number of renewals required to reach a given position, as well as the associated rate functions, is insensitive to the initial conditions, whereas the far tails of the positional distribution retain a dependence on the initial conditions. Our results provide new insights into the rare-event statistics and large deviations of random walks.
\end{abstract}

\maketitle

\section{Introduction}
Anomalous diffusion has emerged as a prominent research topic spanning diverse fields, including biology \cite{Krapf2016Strange}, physics \cite{Metzler2000random,Li2026Biased,Hou2018Biased,Afek2003Colloquium}, chemistry \cite{Aquino2017Chemical}, and mathematics \cite{Kelly2019fractional,Yong2016Backward,Deng2020Modeling}. It describes transport processes that deviate from classical Brownian motion, where the mean squared displacement does not follow the conventional linear dependence on time. Such anomalous behaviors are frequently observed in complex systems characterized by disorder, heterogeneity, or memory effects, including intracellular transport, glassy materials, porous media, and other nonequilibrium systems. Among various theoretical frameworks developed to describe anomalous diffusion, the continuous-time random walk (CTRW) model \cite{Montroll1965Random,Metzler2000random,Li2026Biased,Kutner2017continuous,Vot2020Continuous,Berkowitz2006Modeling,Liu2020Correlated,Hou2018Biased,Cherstvy2019Non,Akimoto2018Ergodicity,Wang2020Fractional,Wang2019Transport,Deng2020Modeling} provides a powerful and versatile approach by incorporating random waiting times and jump lengths, allowing the description of a broad class of anomalous transport phenomena.

In this manuscript, we investigate a case in which the mean squared displacement grows linearly with the observation time $t$, while the corresponding positional distribution is non-Gaussian, contrary to the expectation for normal Brownian diffusion. This phenomenon is sometimes referred to as "anomalous yet Brownian" diffusion \cite{Wang2009Anomalous,Chubynsky2014Diffusing}. In \cite{Pinaki2007Universal}, Chaudhuri, Berthier, and Kob revealed a universal feature in the dynamics of materials approaching the glass transition: the distribution of single-particle displacements develops an exponential tail at large displacements. Subsequently, the displacement probability distribution of particles was also found to exhibit an exponential tail in lipid tubes composed of pure dilauroylphosphatidylcholine bilayers \cite{Wang2009Anomalous}. In particular, the authors noted that although the exponential tail is reminiscent of the slow dynamics typically observed in glassy systems, the underlying dynamics in this case are remarkably rapid. In \cite{Barkai2020Packets}, Barkai and Burov provide, in some sense, a deeper explanation of why the process can exhibit such rapid dynamics: as the number of renewals becomes very large, particles tend to move predominantly in the same direction. Accordingly, the emergence of the exponential decay can be understood in two steps. The first is to characterize the decay of the probability of observing a large number of renewals, while the second is to investigate how this behavior is translated into the decay of the positional distribution. To characterize the statistics of the positional distribution, the rate function of the number of renewals, together with the conditional position given the number of renewals, was employed in \cite{Adrian2021Large}. In particular, several widely used waiting-time distributions were considered.
In addition, a Brownian particle under the influence of a general confining potential that grows with distance as a sublinear power law was investigated. It was found that, for this memoryless model, the observables exhibit stretched-exponential relaxation \cite{Defaveri2024Stretched}. In \cite{defaveri2025diffusion}, Defaveri and Barkai considered a random walk in a random environment on a one-dimensional lattice and used the revival of Laplace's first law of errors to predict the positional distribution of the particles. See also the related recent work discussed in \cite{Hidalgo2020Hitchhiker,hamdi2024laplace}.

In this manuscript, we focus on the far tails of the positional distribution. Our analysis has three aims. First, we extend the results discussed in \cite{Burov2022Exponential,Wang2026Far} by allowing waiting-time densities to diverge as the waiting time approaches zero. Second, we consider a broader range of waiting-time distributions, including distributions with a lower cutoff and one-sided L{\'e}vy distributions, and obtain the most likely number of renewals for the fixed position. Third, we examine the effect of ordinary and equilibrium initial conditions. Although the waiting-time distributions are similar to those considered in \cite{Adrian2021Large}, our analysis focuses on different families of distributions. For example, the generalized power-series expansion covers exponential distributions, Dagum distributions \cite{Barkai2020Packets}, and a special case of the Beta distribution, among others. We show below that the far tails of the positional distribution are sensitive to the behavior of the waiting-time density near the minimum waiting time.

The manuscript is organized as follows. In Sec.~\ref{modelUIIII}, we introduce the CTRW model and define the observables of interest. In Sec.~\ref{SectIIIIW}, we investigate the far tails of the positional distribution for the CTRW model. We consider three types of waiting-time distributions, i.e., a generalized power-series expansion, a cutoff case, and the one-sided L{\'e}vy distribution. In particular, for the generalized power-series case, we analyze the associated rate functions and the ratio of positional distributions. In the following section, we consider the equilibrium CTRW model and investigate the far tails of its positional distribution. Finally, we summarize our main findings and conclusions.

\section{Model}\label{modelUIIII}
We consider the normal CTRW model\cite{Montroll1965Random,Metzler2000random,Li2026Biased,Kutner2017continuous,Vot2020Continuous,Berkowitz2006Modeling,Liu2020Correlated,Burioni2013Rare,Hou2018Biased,Cherstvy2019Non,Akimoto2018Ergodicity,Wang2020Fractional,Wang2019Transport,Deng2020Modeling}, originally introduced by E. Montroll and G. H. Weiss \cite{Montroll1965Random}, in which the waiting times and displacements are assumed to be independent and identically (IID) distributed random variables drawn from the probability density functions $\phi(\tau)$ and $f(x)$, respectively.  Another important quantity is the number of renewals up to time $t$, denoted by $N_t$, which is defined through the relation
\begin{equation}
\tau_1+\tau_2+\cdots+\tau_{N_t}+B_t=t,
\end{equation}
where the waiting times $\tau_i$ are sampled from $\phi(\tau)$ and $B_t$ represents the backward recurrence time \cite{Godreche2001Statistics,Wang2018Renewal}. The position of the particle is then expressed as
\begin{equation}
x(t)=\sum_{i=1}^{N_t}x_i.
\end{equation}
Here the IID displacements $x_i$ are drawn from $f(x)$. For simplicity, we omit the subscript $t$ of $N_t$ in the following discussion. In Laplace space, with $t\to s$, the probability distribution of the number of renewals, denoted by $Q_t(N)$, is given by \cite{Godreche2001Statistics,Wang2018Renewal}
\begin{equation}\label{QtNLapace}
\hat{Q}_{s}(N)=\frac{1-\hat{\phi}(s)}{s}\hat{\phi}(s)^N,
\end{equation}
where $\hat{\phi}(s)$ is the Laplace transform of $\phi(t)$.  Let $P(x,t)$ denote the position distribution of the particle, and let $f(x|N)$ represent the conditional position distribution given that $N$ renewals have occurred. The position distribution can then be written as \cite{Klafter2011First}
\begin{equation}\label{26subInter}
\begin{split}
P(x,t)&=\sum_{N=0}^{\infty}Q_t(N)f(x|N)\\
&=\int_0^\infty Q_t(N)f(x|N)dN .
\end{split}
\end{equation}
Throughout this manuscript, we assume that all particles start from the initial condition $x(t=0)=0$ at $t=0$. In the following analysis, we focus on the statistics of large displacements $x$ in the short-time limit, based on the second line of Eq.~\eqref{26subInter}. In our discussion, we investigate the Gaussian distribution
\begin{equation}\label{AT19SEC2EQs104}
f(x)=\frac{1}{\sqrt{2\pi\sigma^2}}
\exp\left[-\frac{(x-a)^2}{2\sigma^2}\right]
\end{equation}
with mean $a>0$ and variance $\sigma^2$. When $a=0$, Eq.~\eqref{26subInter} reduces to a symmetric positional distribution.
As stated above, the displacements are assumed to be IID random variables. Consequently, after $N$ jumps, the total displacement is given by the sum of $N$ IID Gaussian random variables. The corresponding conditional probability density of the total displacement $x$ given $N$ renewals is therefore also Gaussian, with mean $aN$ and variance $\sigma^2N$
\begin{equation}\label{AT19SEC2EQs105}
f(x|N)=\frac{1}{\sqrt{2\pi\sigma^2N}}
\exp\left[-\frac{(x-aN)^2}{2\sigma^2N}\right].
\end{equation}
Below, we investigate the far tails of the positional distribution for the ordinary and equilibrium CTRW models, respectively.

\section{Far tails of the positional distribution for the Normal continuous-time random walk}\label{SectIIIIW}
Motivated by the work presented in \cite{Wang2026Statistics}, we consider three families of waiting-time PDFs: a generalized power-series expansion, a power-series expansion with a cutoff, and the one-sided L{\'e}vy distribution. We will show that the far-tail behavior of the positional distribution is largely determined by the small-$\tau$ expansion of the waiting-time distribution. We then employ the saddle-point method to investigate the positional distribution, deriving both  the ratio of the positional distributions with and without bias, and the associated rate functions.

\subsection{Generalized power-series expansion}
\subsubsection{Far tail of the positional distribution}
\begin{figure}[tp!]
 \centering
 \includegraphics[width=0.48\textwidth]{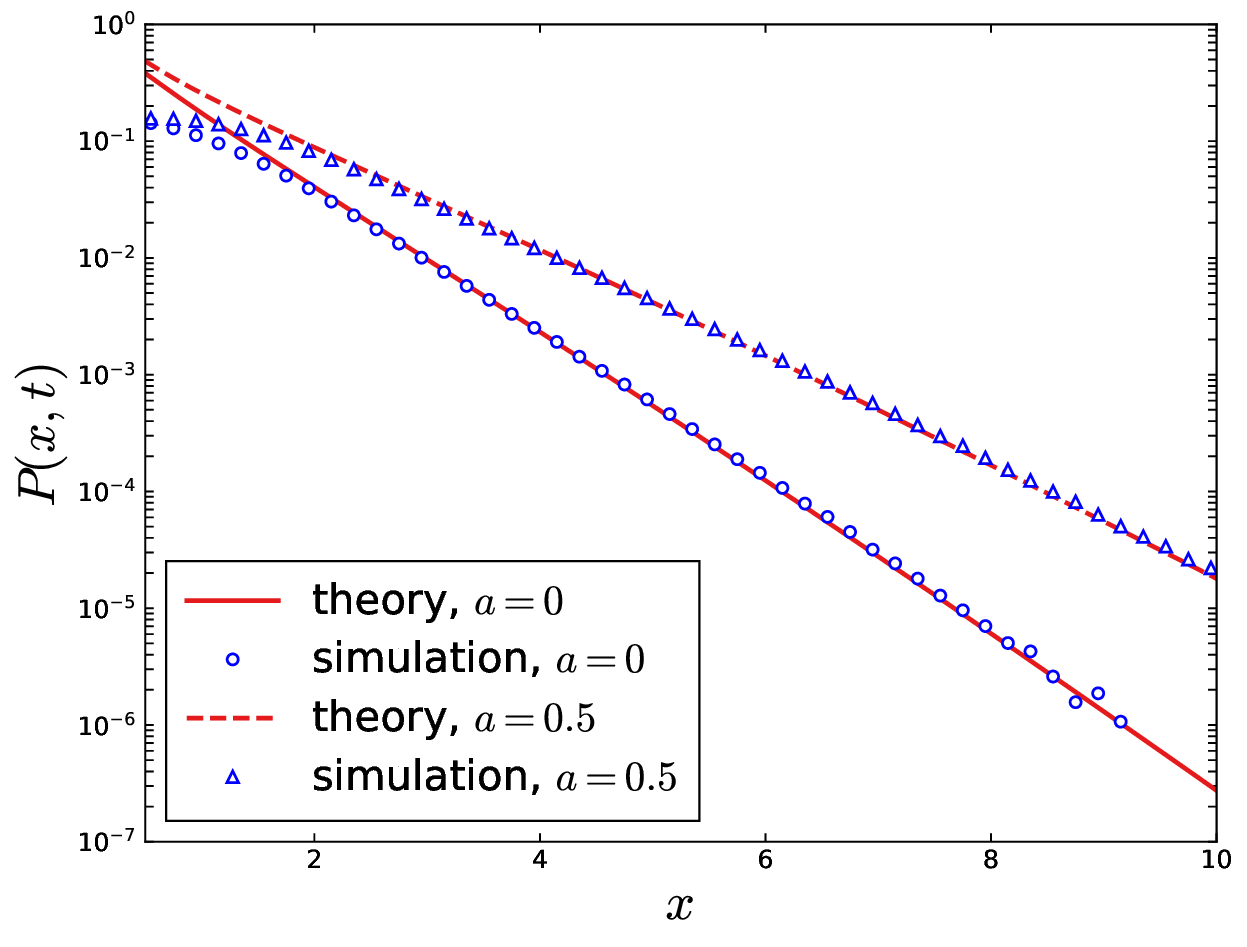}\vspace{-0.3cm}
 \caption{The exponential decay of the positional distribution for Mittag--Leffler waiting times combined with Gaussian displacements is investigated. In our simulations, the waiting-time distribution is given by
$\phi(\tau)=\tau^{\lambda-1}E_{\lambda,\lambda}(-\tau^\lambda)$,
which has both infinite mean and variance. For small $\tau$, one has the expansion
$\phi(\tau) \sim \tau^{\lambda-1}/\Gamma(\lambda) -\tau^{2\lambda-1}/\Gamma(2\lambda)$.
In the special case $\lambda=1/2$, the corresponding coefficients are $C_{\alpha}=1/\sqrt{\pi}$ with $\alpha=-1/2$, and $C_{\beta}=-1$ with $\beta=0$. For the displacement statistics, we set $\sigma=1$ and $a=0$. The solid line represents the theoretical prediction from Eq.~\eqref{26Sec2108}, while the symbols correspond to the simulation results.
}\label{mlfPxta0}
\end{figure}

We first consider the case where $\phi(\tau)$ is characterized by a generalized power-series expansion in the vicinity of $\tau=0$
\begin{equation}\label{26Sec2101}
\phi(\tau)=C_{\alpha} \tau^{\alpha}+C_{\beta} \tau^{\beta}+\cdots, \quad(\tau \rightarrow 0)
\end{equation}
with $\beta>\alpha>-1$. If $-1<\alpha<0$, Eq.~\eqref{26Sec2101} tends to infinity for $\tau\to 0$. The special case in which $\alpha$ and $\beta$ are non-negative integers was previously studied in Ref.~\cite{Burov2022Exponential,Wang2026Far} for the CTRW model. Here, we generalize these results to a wider setting. For Eq.~\eqref{26Sec2101}, in the context of a renewal process, the number of renewals follows \cite{Wang2026Statistics}
\begin{equation}\label{26Sec2102}
Q_{t}(N) \sim \frac{\left(\left[C_{\alpha} \Gamma(1+\alpha)\right]^{1 /(1+\alpha)} t\right)^{N(1+\alpha)}}{\Gamma((\alpha+1) N+1)} \exp \left(\frac{N^{1+\alpha-\beta}}{(1+\alpha)^{\beta-\alpha}} \frac{C_{\beta} \Gamma(1+\beta)}{C_{\alpha} \Gamma(1+\alpha)} t^{\beta-\alpha}\right),
\end{equation}
illustrating rare events of the distribution. This equation is valid in the large-$N$, short-observation-time $t$ regime. By rewriting Eq.~\eqref{26Sec2102}, one can obtain the nearly exponential decay \cite{Wang2026Statistics}.
Since the displacements have finite mean and variance, large positional fluctuations are dominated by trajectories with a large number of renewals.
Thus, substituting Eq.~\eqref{26Sec2102} into the second line of Eq.~\eqref{26subInter}, we have
\begin{equation}\label{26Sec2103}
P(x,t)\sim\int_0^\infty \exp\left(
\underbrace{\begin{gathered}
(\alpha+1)N\ln\left(\frac{zet}{N}\right)
+\frac{C_{\beta}\Gamma(\beta+1)N^{\alpha-\beta+1}t^{\beta-\alpha}}{(\alpha+1)^{\beta-\alpha}C_{\alpha}\Gamma(\alpha+1)}
\\
-\frac{(x-aN)^2}{2N\sigma^2}
-\frac12\ln\left(4\pi^2(\alpha+1)N^2\sigma^2\right)
\end{gathered}}_{\chi(N)}\right)dN
\end{equation}
with $z=(C_{\alpha} \Gamma(\alpha+1))^{1/(\alpha+1)}/(\alpha+1)$.
To derive the asymptotic solution, we apply the saddle-point method. The saddle point is determined by taking the derivative with respect to $N$ and imposing the stationary condition
\begin{align*}
\chi^{\prime}(N)\sim{}&\frac{x^2}{2\sigma^2N^2}-\frac{a^2}{2\sigma^2}
-(\alpha+1)\ln\left(\frac{N}{zt}\right)-\frac1N
\\
&+\frac{C_{\beta}\Gamma(\beta+1)(\alpha-\beta+1)}{C_{\alpha}\Gamma(\alpha+1)(\alpha+1)^{\beta-\alpha}}
\left(\frac{t}{N}\right)^{\beta-\alpha}=0,
\end{align*}
In the limit of large $N$, the asymptotic solution of the above equation yields
\begin{equation}\label{NxingGeneralexpansion}
N^{*} \sim \frac{|x|}{\sigma \sqrt{(\alpha+1) W_{0}\left[\frac{\exp \left(\frac{a^{2}}{(\alpha+1) \sigma^{2}}\right)}{(\alpha+1) z^{2} \sigma^{2}} \frac{x^{2}}{t^{2}}\right]}}.
\end{equation}
Here, $W_{0}(z)$ is the principal branch of the Lambert $W$ function \cite{CorlessOn1996,Zarfaty2016Statistics,Zarfaty2018Dispersion}, defined through $w\exp(w)=z$ with the solution $w=W_{0}(z)$. Its asymptotic forms are $W_{0}(z)\sim \ln z$ for $z\gg e$ and $W_{0}(z)\sim z$ as $z\to 0$. Comparison with the result reported in \cite{Wang2026Far} shows that the saddle point $N^{*}$ is essentially unaffected by the correction term in the exponent of Eq.~\eqref{26Sec2102}. Using the saddle point method, the asymptotic tails  of the position follows
\begin{equation}\label{saddlePointMethod}
P(x, t) \sim \frac{\sqrt{2 \pi} \exp \left(\chi\left(N^{*}\right)\right)}{\sqrt{\left|\chi^{\prime \prime}\left(N^{*}\right)\right|}}
\end{equation}
with
\begin{equation*}
\begin{aligned}
\chi^{\prime\prime}(N)
&=-\frac{x^{2}}{\sigma^{2}N^{3}}-\frac{\alpha+1}{N}\quad-\frac{C_{\beta}\Gamma(\beta+1)(\alpha-\beta+1)t^{\beta-\alpha}(\beta-\alpha)}
{C_{\alpha}\Gamma(\alpha+1)(\alpha+1)^{\beta-\alpha}N^{\beta-\alpha+1}}
+\frac{1}{N^{2}}.
\end{aligned}
\end{equation*}
In the limit of the large $x$, we have
\begin{equation}\label{26Sec2108}
\begin{aligned}
P(x,t)\sim{}&
\frac{
W_0^{1/4}\left(g_a\frac{x^2}{t^2}\right)
}{
\sqrt{
2\pi\sigma(1+\alpha)^{3/2}|x|
}
}
\Bigg[
W_0\left(g_a\frac{x^2}{t^2}\right)
+1
+
R_1
\left(\frac{t}{|x|}\right)^{\beta-\alpha}
W_0^{\frac{\beta-\alpha}{2}}
\left(g_a\frac{x^2}{t^2}\right)
\Bigg]^{-1/2}
\\
&\times
\exp\Bigg[
-|x|G_a\left(\frac{x}{t}\right)
+\frac{a}{\sigma^2}x
+\sqrt{1+\alpha}\,R_2 t
\left(\frac{t\sigma}{|x|}\right)^{\beta-\alpha-1}
W_0^{\frac{\beta-\alpha-1}{2}}
\left(g_a\frac{x^2}{t^2}\right)
\Bigg]
\end{aligned}
\end{equation}
with
\begin{equation}
R_{1}=\frac{C_{\beta} \Gamma(\beta+1)(\alpha-\beta+1)(\beta-\alpha) \sigma^{\beta-\alpha} }{C_{\alpha} \Gamma(\alpha+2)(\alpha+1)^{\frac{\beta-\alpha}{2}}},
\end{equation}

\begin{equation}
R_2=\frac{C_{\beta} \Gamma(\beta+1)}{C_{\alpha} \Gamma(\alpha+2)(\alpha+1)^{\frac{\beta-\alpha}{ 2}}},
\end{equation}
\begin{equation}\label{GGGa}
G_a(l)=\frac{\sqrt{\alpha+1}}{\sigma}\left(\sqrt{W_{0}\left(g l^{2}\right)}-\frac{1}{ \sqrt{W_{0}\left(g l^{2}\right)}}\right),
\end{equation}
and
\begin{equation}\label{ggga}
g_a=\frac{\exp \left(\frac{a^{2}}{ (\alpha+1) \sigma^{2}}\right)}{(\alpha+1) \sigma^{2} z^{2}}.
\end{equation}
For simplicity, the subscript $a$ is omitted henceforth, except when $a$ takes a special value, such as $a=0$, in which case we explicitly write $G_0(l)$ instead of $G_a(l)$. The same notation convention is adopted for $g_a$. See Fig.~\ref{mlfPxta0} for a comparison with the simulations.
For large $x$, the leading term of Eq.~\eqref{26Sec2108} decays as
\begin{equation}\label{AsyNorPpxt}
P(x,t)\sim \exp\left(-|x|\frac{\sqrt{\alpha+1}}{\sigma}\sqrt{W_{0}\left(g l^{2}\right)}+\frac{a}{\sigma^2}x\right)
\end{equation}
following the nearly exponential decay. This indicates that, whenever the waiting-time distribution follows the small $\tau$ behavior specified in Eq.~\eqref{26Sec2101}, the far tails of the positional distribution decay exponentially.

\subsubsection{Ratio of the positional distribution with and without bias}
\begin{figure}[tp!]
 \centering
 \includegraphics[width=0.48\textwidth]{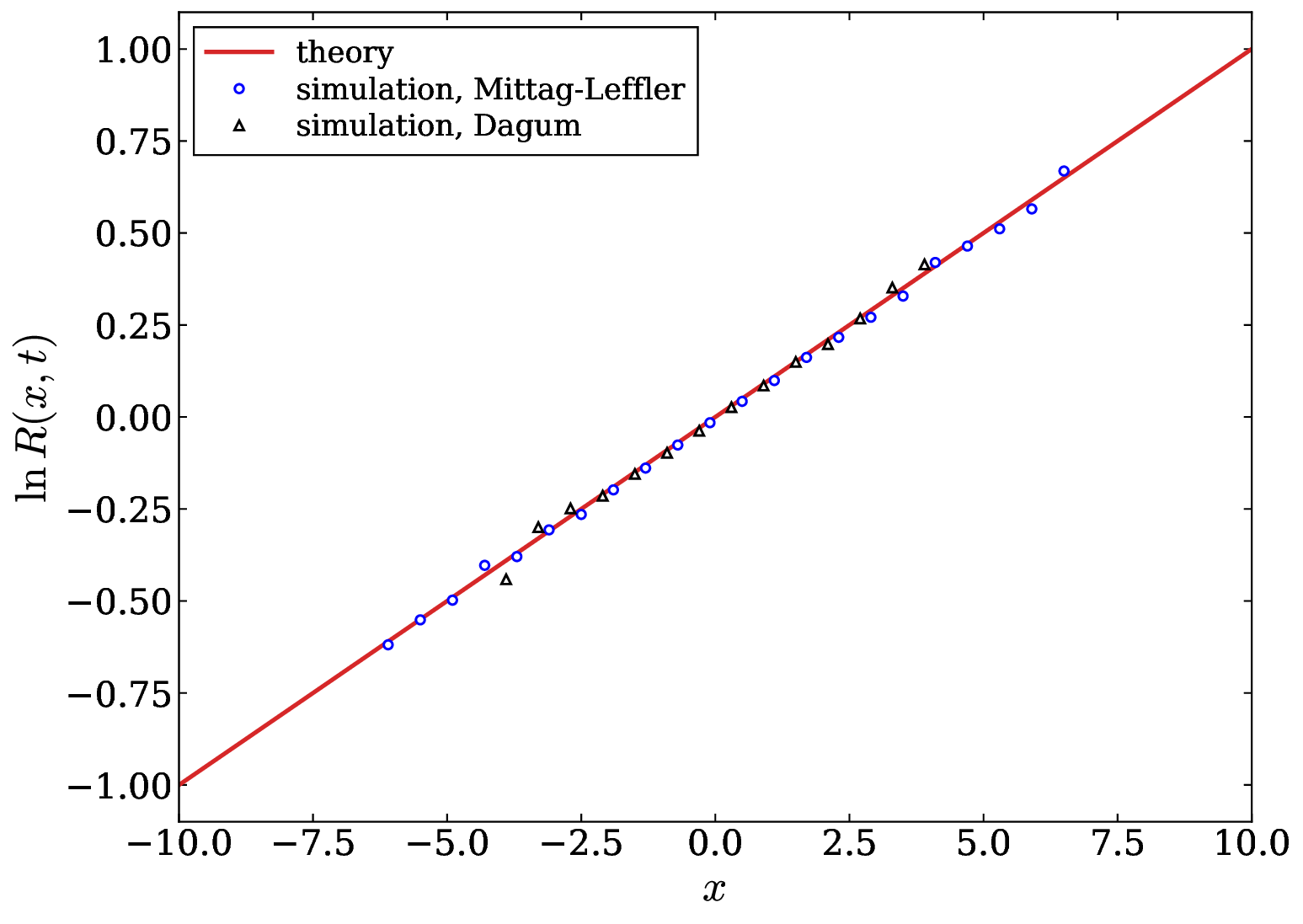}\vspace{-0.3cm}
 \caption{Plot of $\ln\mathcal{R}(x,t)$, showing its linear dependence on the position $x$. The solid line represents the theoretical prediction given by Eq.~\eqref{ratioOfPositionalBiased2}, while the symbols denote simulation results for the Mittag-Leffler and Dagum distributions. The theoretical prediction agrees perfectly with the simulation results. The parameters are $\phi(\tau)=\tau^{\lambda-1}E_{\lambda,\lambda}(-\tau^\lambda)$~(Mittag-Leffler distribution) with $\lambda=1/2$, $\phi(\tau)=1/(1+\tau)^2$ (Dagum distribution), $a=0.1$, and $\sigma=1$.
}\label{RatioOfMlfandDagum}
\end{figure}

Below we investigate the effect of the bias by analyzing the tail statistics. More specifically, we consider the ratio between the tails of the biased and unbiased distributions, which is defined as
\begin{equation}\label{RatioRRR}
\mathcal{R}(x, t)=\frac{P(x, t)|_{a \neq 0}}{P(x, t)|_{a=0}}.
\end{equation}
Substituting Eq.~\eqref{26Sec2108} into \eqref{RatioRRR}, we have
\begin{equation*}
\mathcal{R}(x,t)\sim\exp\left[\begin{aligned}
&\frac{ax}{\sigma^2}-x\operatorname{sign}(x)
\left(G_a\left(\frac{x}{t}\right)-G_0\left(\frac{x}{t}\right)\right)
\\
&+\sqrt{1+\alpha}\,R_2t\left(\frac{t\sigma}{|x|}\right)^{\beta-\alpha-1}
\Bigg(\left(W_0\left(\frac{gx^2}{t^2}\right)\right)^{\frac{\beta-\alpha-1}{2}}
\\
&\qquad\qquad-\left(W_0\left(\frac{g_0x^2}{t^2}\right)\right)^{\frac{\beta-\alpha-1}{2}}\Bigg)
\end{aligned}\right].
\end{equation*}
Here, the sign function (or the signum function), denoted by ${\rm sign}(x)$, specifies the sign of a real number. It takes the value $1$ for positive numbers, $-1$ for negative numbers, and $0$ when the argument is zero.  The functions $G_a(y)$ and $g_a(y)$ are defined by Eqs.~\eqref{GGGa} and \eqref{ggga}, respectively.
Rewriting the above equation, we obtain
\begin{equation}\label{ratioOfPositionalBiased}
\begin{aligned}
\ln(\mathcal{R}(x,t))\sim{}&\frac{ax}{\sigma^2}-x\operatorname{sign}(x)
\left(G_a\left(\frac{x}{t}\right)-G_0\left(\frac{x}{t}\right)\right)
\\
&+\sqrt{1+\alpha}\,R_2t\left(\frac{t\sigma}{|x|}\right)^{\beta-\alpha-1}
\Bigg(\left(W_0\left(\frac{gx^2}{t^2}\right)\right)^{\frac{\beta-\alpha-1}{2}}
\\
&\qquad\qquad-\left(W_0\left(\frac{g_0x^2}{t^2}\right)\right)^{\frac{\beta-\alpha-1}{2}}\Bigg)
\end{aligned}
\end{equation}
Recall that both $G_a(y)$ and $g_a(y)$ are even functions of $a$. Expanding them in a Taylor series around $a=0$, we obtain the following simple asymptotic expression in the small-$a$ limit
\begin{equation}\label{ratioOfPositionalBiased2}
\ln(\mathcal{R}(x,t))\sim \frac{a x}{\sigma^2}.
\end{equation}
See the plot in Fig.~\ref{RatioOfMlfandDagum}. The theoretical prediction shows that the logarithm of the ratio depends linearly on the position and is independent of the statistical properties of the waiting-time distribution. This feature is particularly valuable experimentally, as it allows the prediction to be tested without measuring or characterizing the waiting-time distribution of the particles.  Note that Eq.~\eqref{ratioOfPositionalBiased2} breaks down in the strong-bias regime, while Eq.~\eqref{ratioOfPositionalBiased} remains valid for bias of any strength.

\begin{figure}[tp!]
 \centering
 \includegraphics[width=0.48\textwidth]{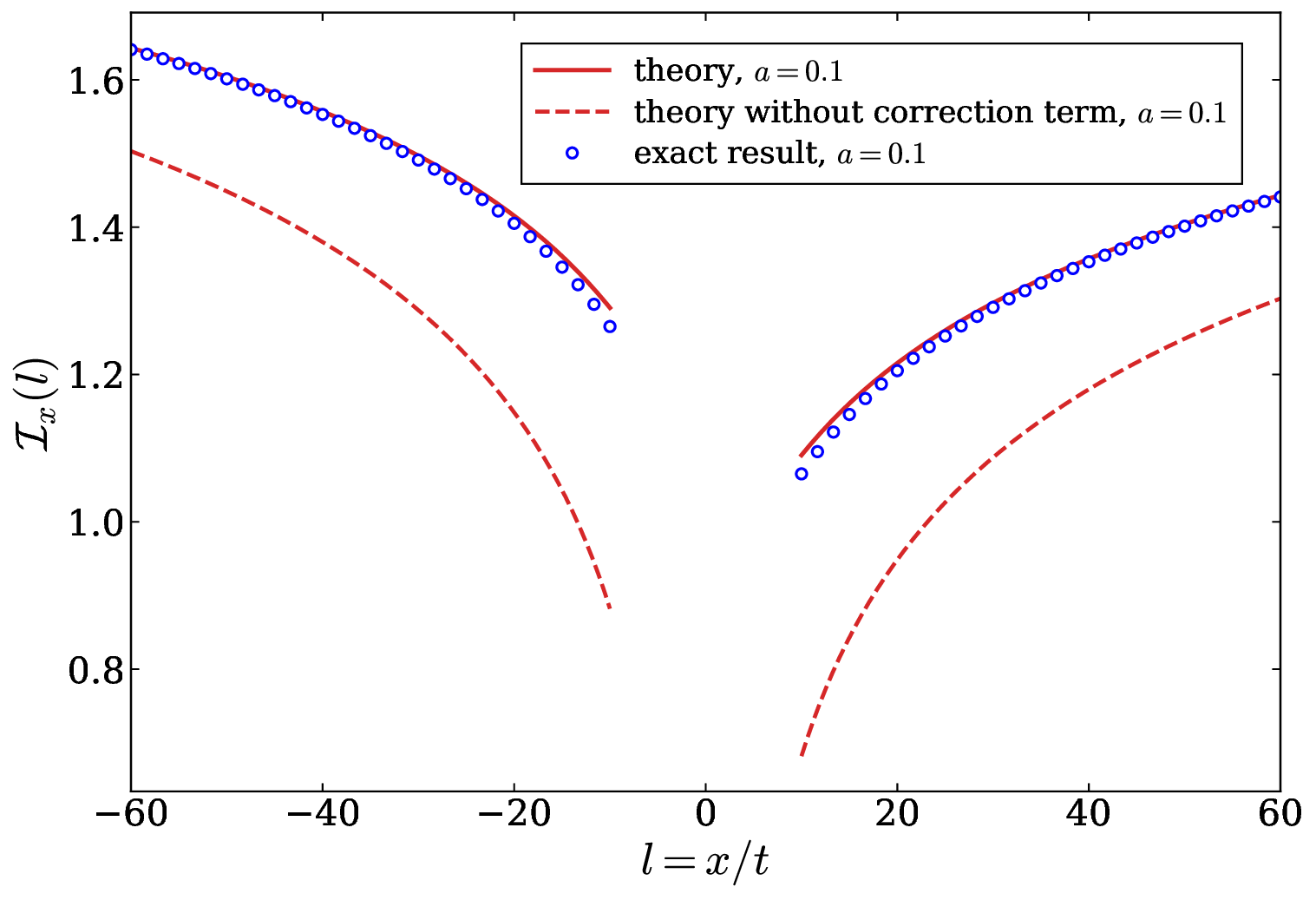}\vspace{-0.3cm}
 \caption{Rate function for the Mittag-Leffler waiting-time distribution. The solid line shows the theoretical prediction given by Eq.~\eqref{rateFunction11}, while the dashed line shows the prediction of Eq.~\eqref{rateFunction11} without the last term (correction term). We choose $\phi(\tau)=\tau^{\lambda-1}E_{\lambda,\lambda}(-\tau^\lambda)$ with $\lambda=1/2$, $a=0.1$, and $\sigma=1$.
}\label{RateMlf}
\end{figure}

\subsubsection{Rate function}

We next consider the rate function, which plays a fundamental role in large-deviation theory by quantifying the probabilities of rare fluctuations \cite{Touchette2009large,Barkai2020Packets,Wang2020Large,Adrian2021Large,Wang2026Statistics}. The rate function is also known as the Cram{\'e}r function, named after the Swedish probabilist Harald Cram{\'e}r.
Below, we derive the positional and temporal rate functions associated with the asymptotic limits $x\to\infty$ and $t\to\infty$, respectively.
 The derivation is based on determining the positional dependent function $\mathcal{I}_x(l)$ that satisfies
\begin{equation}
\lim _{x \rightarrow \infty} \frac{\ln (P(x, t))}{-|x|}=\mathcal{I}_{x}\left(l=\frac{x}{t}\right).
\end{equation}
The positional rate function can be obtained through the following procedure. We first take the logarithm of Eq.~\eqref{26Sec2108}, divide it by $-|x|$, and then take the limit $|x|\to\infty$. This yields the desired positional rate function
\begin{equation}\label{rateFunction11}
\mathcal{I}_{x}(l)=G(l)-\frac{a}{\sigma^{2}} \operatorname{sign}(l)-\frac{C_{\beta} \Gamma(\beta+1) \cdot \sigma^{\beta-\alpha-1} \cdot\left[W_{0}\left(g l^{2}\right)\right]^{\frac{\beta-\alpha-1}{2}}}{C_{\alpha} \Gamma(\alpha+1) \cdot(\alpha+1)^{\frac{\beta-\alpha+1}{2}} \cdot|l|^{\beta-\alpha}}.
\end{equation}
When $\beta=\alpha+1$ and $\alpha$ is a non-negative integer, the last term on the right-hand side of Eq.~\eqref{rateFunction11} reduces to $C_{\alpha+1}/C_{\alpha}|l|$, as discussed in \cite{Wang2026Far}. However, in the general case, this correction term, i.e., the last term on the right-hand side of Eq.~\eqref{rateFunction11}, is important, as illustrated in Fig.~\ref{RateMlf}.
It can be seen that the dominate  term of  Eq.~\eqref{rateFunction11} is
\begin{equation*}
I_{x} \sim G(l) \sim \frac{\sqrt{\alpha+1}}{\sigma} \cdot\left(\sqrt{\ln \left(g l^{2}\right)}-\frac{1}{\sqrt{\ln \left(g l^{2}\right)}}\right) \sim \frac{\sqrt{\alpha+1}}{\sigma} \sqrt{\ln \left(g l^{2}\right)},
\end{equation*}
which is responsible for the nearly exponential decay of the positional distribution in the limit of $x/t\to\infty$. Similarly, the time rate function can be achieved by using
\begin{equation}
\lim_{t \rightarrow \infty}\frac{\ln (P(x, t))}{-t}=\mathcal{I}_{t}\left(l=\frac{x}{t}\right)
\end{equation}
with
$\mathcal{I}_{t}(l)=\operatorname{sign}(l)l\mathcal{I}_{x}(l)$.

\subsection{Waiting time distribution with a cutoff}
\begin{figure}[tp!]
 \centering
 \includegraphics[width=0.48\textwidth]{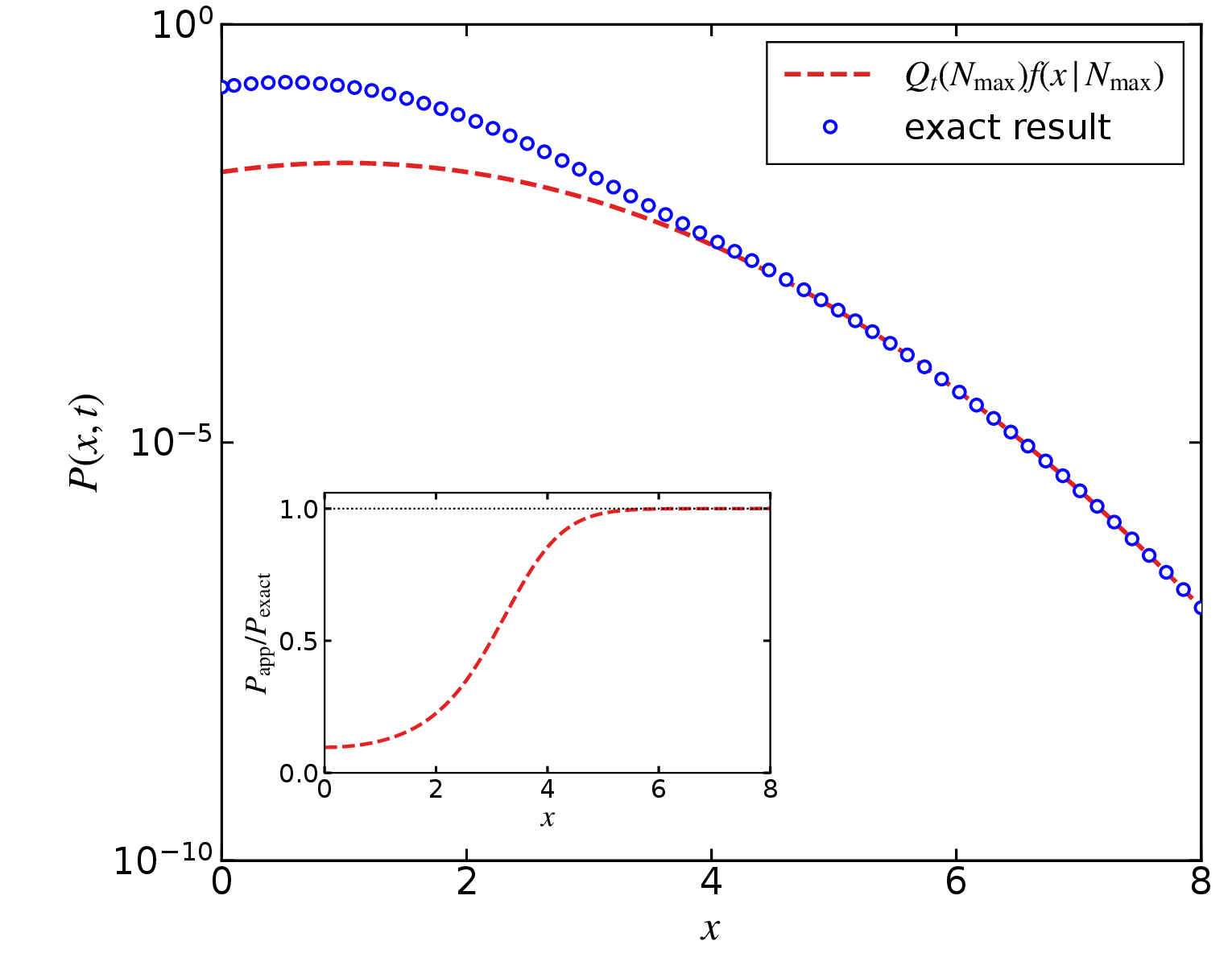}\vspace{-0.3cm}
 \caption{The far tails of the positional distribution for a waiting-time distribution with a cutoff. The dashed line represents the Gaussian decay predicted by Eq.~\eqref{shwh0}, while the symbols denote the exact results obtained by evaluating the infinite sum in Eq.~\eqref{26subInter}. The inset shows the ratio of the prediction in Eq.~\eqref{shwh0} to the exact results from Eq.~\eqref{26subInter}, demonstrating excellent agreement for the tails.
We choose $\phi(\tau)=1/(1+\tau-\tau_0)^2$,  $t=0.5$, $\tau_0=0.2$, $a=0$, $\sigma^2=1$, and $N_{\max}=2$.
}\label{ordinary_dagum_t0p5_tau0p2_a0p5}
\end{figure}

We next examine the truncated waiting-time case, where the distribution is cut off at $\tau_0$, meaning no waiting times fail to meet this threshold. For this case, its expansion follows
\begin{equation}\label{cutoff101}
\phi(\tau)\sim \left(C_\alpha(\tau-\tau_0)^\alpha+C_\beta(\tau-\tau_0)^\beta+\cdots \right)\Theta(\tau-\tau_0),
\end{equation}
Clearly, $\phi(\tau) = 0$ for $\tau < \tau_0$. Compared with Eq.~\eqref{26Sec2101}, the current expression is obtained by shifting the parameter $\tau_0$ with respect to that reference. Under this condition, it follows from Eq.~\eqref{QtNLapace} that the asymptotic tail of the renewal number is given by \cite{Wang2026Far}
\begin{equation}\label{cutoffQtn}
\begin{aligned}
Q_t(N)\sim{}&\frac{\left[\left(C_{\alpha}\Gamma(1+\alpha)\right)^{\frac1{(1+\alpha)}}\left(t-N\tau_0\right)\right]^{N(1+\alpha)}}{\Gamma((\alpha+1)N+1)}
\\
&\cdot\exp\left(\frac{C_{\beta}\Gamma(1+\beta)}{C_{\alpha}\Gamma(1+\alpha)}
\frac{N\left(t-\tau_0N\right)^{\beta-\alpha}}{((\alpha+1)N+1)^{\beta-\alpha}}\right)
\times\Theta\left(t-\tau_0N\right).
\end{aligned}
\end{equation}
Consequently, the above equation implies that the far tail of the position takes the form
\begin{equation}\label{26Sec2303}
P(x,t)\sim\int_0^{\lfloor t/\tau_0\rfloor}\exp\left(
\underbrace{\begin{gathered}
(\alpha+1)N\ln\left(\frac{ze(t-N\tau_0)}{N}\right)
\\
+\frac{C_{\beta}\Gamma(\beta+1)N(t-\tau_0N)^{\beta-\alpha}}{C_{\alpha}\Gamma(\alpha+1)((1+\alpha)N+1)^{\beta-\alpha}}
\\
-\frac{(x-aN)^2}{2N\sigma^2}
-\frac12\ln\left(4\pi^2(\alpha+1)N^2\sigma^2\right)
\end{gathered}}_{\chi(N)}\right)dN.
\end{equation}
Because of the cutoff in the waiting-time distribution, the summation/integral over $N$ is restricted to the range $0 \le N \le t/\tau_0$, in contrast to Eq.~\eqref{26Sec2103}. The strong correlation between $N$ and $t$ prevents the saddle-point equation from being solved in closed form. However, in this case, we can use the maximum number of renewals ($N_{\max}=\lfloor  t/\tau_0\rfloor$) to approximate the far tails of the positional distribution
\begin{equation}\label{shwh0}
    P(x,t)\sim\frac{ Q_t(\lfloor  t/\tau_0\rfloor)}{\sqrt{2\pi\sigma^2\lfloor  t/\tau_0\rfloor}}
\exp\left[-\frac{(x-a\lfloor  t/\tau_0\rfloor)^2}{2\sigma^2\lfloor  t/\tau_0\rfloor}\right],~~~ -\infty<x<+\infty.
\end{equation}
See the plot in Fig.~\ref{ordinary_dagum_t0p5_tau0p2_a0p5}. For a waiting-time distribution with a cutoff, the number of renewals has a finite upper bound, whereas the positional distribution has no cutoff and instead exhibits a Gaussian decay in its far tails. See Ref.~\cite{Adrian2021Large} for a discussion of the cutoff case, in which a numerical solution is employed.

\subsection{The one-sided L{\'e}vy distribution}
Our third example is the widely used one-sided L{\'e}vy distribution, which is notable for having both infinite mean and infinite variance \cite{Schneider1986Stable,Zolotarev1986One}. In Laplace space, it takes the simple form $\mathcal{L}[\phi(\tau)] = \exp(-s^\alpha)$.
For small $\tau$, the waiting-time density has the asymptotic form
\[
\begin{aligned}
\phi(\tau)\sim{}&
\frac{\alpha^{\frac{1}{2(1-\alpha)}}}{\sqrt{2\pi(1-\alpha)}}
\tau^{-\frac{2-\alpha}{2(1-\alpha)}}\exp\left[-(1-\alpha)\alpha^{\frac{\alpha}{1-\alpha}}
\tau^{-\frac{\alpha}{1-\alpha}}\right].
\end{aligned}
\]
Substituting $\widehat{\phi}(s)=\exp(-s^\alpha)$ into Eq.~\eqref{QtNLapace}, the asymptotic distribution of the number of renewals is \cite{Wang2026Far}
\begin{equation}
\begin{aligned}
Q_t(N)\sim{}&\frac{1}{\sqrt{2\pi(1-\alpha)}}
\left(\frac{t^\alpha}{\alpha N}\right)^{\frac{1}{2(1-\alpha)}}
\exp\left[-(1-\alpha)\alpha^{\frac{\alpha}{1-\alpha}}
\left(\frac{N}{t^\alpha}\right)^{\frac{1}{1-\alpha}}\right],
\end{aligned}
\end{equation}
which decays at a rate faster than that of an exponential function.
Similar to Eq.~\eqref{26Sec2103}, the positional distribution leads to
\begin{equation*}
P(x,t)\sim\int_0^\infty \exp\left(
\begin{gathered}
-\alpha^{\frac{\alpha}{1-\alpha}}(1-\alpha)
\left(\frac{N}{t^\alpha}\right)^{\frac{1}{1-\alpha}}
-\frac{1}{2(1-\alpha)}\ln\left(\frac{N}{t^\alpha}\right)
\\
-\frac12\ln\left[2\pi(1-\alpha)\alpha^{\frac{1}{1-\alpha}}\right]
-\frac12\ln(2\pi\sigma^2N)
-\frac{(x-aN)^2}{2\sigma^2N}
\end{gathered}
\right)\,dN.
\end{equation*}
We use the saddle-point method to calculate the far-tail behavior of the positional distribution. The saddle point $N^*$ is determined by setting the derivative of the exponent to zero:
\begin{equation}\label{NxingOnedeid}
\chi^{\prime}(N)\sim -\alpha^{\frac{\alpha}{1-\alpha}} \cdot\left(t^{\alpha}\right)^{-\frac{1}{1-\alpha}} \cdot N^{\frac{\alpha}{1-\alpha}}-\frac{1}{2(1-\alpha)N}-\frac{1}{2 N}+\frac{x^{2}-a^{2} N^{2}}{2 \sigma^{2} N^{2}}=0.
\end{equation}
In the presence of a general bias, the above equation admits no closed-form solution in general. Thus, we begin by examining a bias free case. From the above equation, the asymptotic solution for large $N$ is obtained as
\begin{equation}\label{nixingOneSide}
N^{*} \sim\left( \frac{x^{2} t^{\frac{\alpha}{1-\alpha}}}{2\sigma^{2} \alpha^{\frac{\alpha}{1-\alpha}}} \right)^{\frac{1-\alpha}{2-\alpha}}.
\end{equation}
Equation~\eqref{nixingOneSide} gives $N^*\propto |x|^{2(1-\alpha)/(2-\alpha)}t^{\alpha/(2-\alpha)}$. For $0<\alpha<1$, the power $2(1-\alpha)/(2-\alpha)$ lies between zero and one. Thus, at fixed $t$, the most likely renewal count grows sublinearly with $|x|$ and tends to infinity in the far tail. The combination $|x|t^{\alpha/[2(1-\alpha)]}$ describes the scaling of $N^*$; it is not be identified with the scaling variable of the positional distribution. As shown in Eq.~\eqref{LevyOnePxt} below, the leading decay exponent of the propagator is controlled by $|x|/t^{\alpha/2}$, while its prefactor has additional time dependence. For a discussion of the scaling of typical displacement fluctuations, see Ref.~\cite{Schneider1989Fractional}; this scaling differs from the one considered here.

\begin{figure}[tp!]
 \centering
 \includegraphics[width=0.48\textwidth]{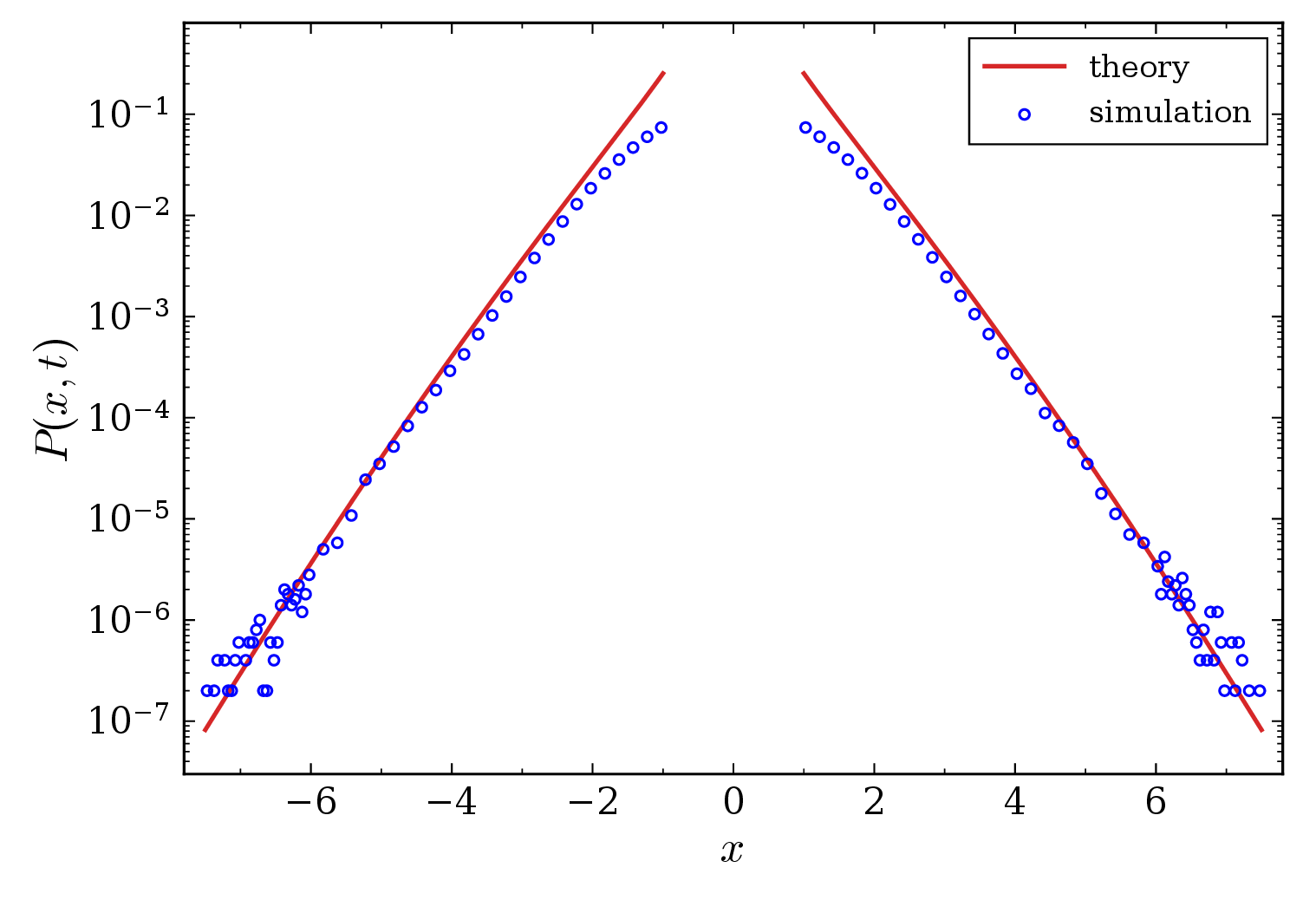}\vspace{-0.3cm}
 \caption{Far tails of the positional distribution for the one-sided L{\'e}vy waiting-time distribution, $\phi(\tau)=\exp\left(-1/(4\tau)\right)/(2\sqrt{\pi}\tau^{3/2})$, corresponding to $\alpha=1/2$. Here, we choose $t=0.5$, $a=0$, and $\sigma=1$. The solid line shows Eq.~\eqref{LevyOnePxt}, while the symbols represent simulation results.
}\label{levy_eq30_mc_signed_alpha0p5_t0p5_bin005}
\end{figure}

From Eq.~\eqref{saddlePointMethod}, the far tail of the positional distribution with the one-sided L{\'e}vy distribution follows
\begin{equation}\label{LevyOnePxt}
\begin{aligned}
P(x,t)\sim{}&\frac1{\sqrt{2\pi(2-\alpha)}}
\left(\frac2{\alpha^2}\right)^{\frac{1+\alpha}{2(2-\alpha)}}
\sigma^{\frac{2\alpha-1}{2-\alpha}}t^{\frac{3\alpha}{2(2-\alpha)}}|x|^{-\frac{1+\alpha}{2-\alpha}}
\\
&\times\exp\left[-(2-\alpha)\left(\frac{\alpha^\alpha}{2\sigma^2t^\alpha}\right)^{\frac1{2-\alpha}}|x|^{\frac2{2-\alpha}}\right].
\end{aligned}
\end{equation}
As shown in Fig.~\eqref{levy_eq30_mc_signed_alpha0p5_t0p5_bin005}, the tails decay faster than exponentially.
It is easy to check that our leading term is consistent with the results discussed in Ref.~\cite{Adrian2021Large}. The advantage of Eq.~\eqref{LevyOnePxt} is that it also allows us to determine the perfector.

\begin{remark}
For the biased case, it is hard to obtain $N^{*}$, except for some special case, for example $\alpha=1/2.$ For this case, using Cardano formula,
\begin{equation*}
N^{*}\sim (A+B)^{\frac{1}{3}}+(A-B)^{\frac{1}{3}}-\frac{a^{2} t}{3 \sigma^{2}} 
\end{equation*}
with $A=x^{2} t /\left[2 \sigma^{2}\right]-a^{6} t^{3} /\left[27 \sigma^{6}\right]$ and $B=\sqrt{A^{2}-a^{12} t^{6} /\left[729 \sigma^{12}\right]}$. Once $N^{*}$ is obtained, the far tail is obtained from Eq.~\eqref{saddlePointMethod}. Besides, when the bias of the system is weak, the perturbation approximation can be used. Specifically, we write the biased saddle point as $N^{*}=N_0^{*}+\delta N$, where $N_0^{*}$ is the saddle point in the unbiased case and $\delta N$ is a small correction induced by the bias. The correction $\delta N$ can then be determined perturbatively by expanding the saddle-point equation around $N_0^{*}$.
\end{remark}

\section{Far Tails of the Positional Distribution for Equilibrium CTRWs}
For the last section, we consider the normal CTRW model, now we turn to the equilibrium CTRW. As mentioned above, in the equilibrium CTRW model, the process is assumed to have started long before the observation begins. More precisely, the first waiting time is drawn from the equilibrium waiting-time distribution \cite{Barkai2003Aging,Schulz2014Aging,Wang2019Transport}
\begin{equation}\label{eqCTRW}
h_{\mathrm{eq}}(\tau)=\frac{\displaystyle\int_\tau^\infty \phi(y)\,dy}{\langle\tau\rangle},
\end{equation}
where $\langle\tau\rangle$ denotes the mean waiting time. After the first waiting time, the particle makes a jump drawn from the jump-length distribution $f(x)$. Subsequently, the particle waits for another waiting time $\tau_2$ drawn from $\phi(\tau)$, followed by another jump drawn from $f(x)$, and this procedure is repeated.
Thus, compared with the  normal CTRW, the only difference lies in the distribution of the first waiting time: it is drawn from $h_{\mathrm{eq}}(\tau)$ in the equilibrium CTRW, rather than from $\phi(\tau)$. Note that the construction of an equilibrium process requires a finite mean waiting time, $\langle\tau\rangle<\infty$. This condition is therefore necessary for the equilibrium process defined in Eq.~\eqref{eqCTRW} to be well defined. Using the aging renewals process,
in Laplace space, the distribution of the number of renewals behave as \cite{Barkai2003Aging,Schulz2014Aging,Wang2019Transport}
\begin{equation}\label{equilibriumQTS}
\hat{Q}_{s}^{\mathrm{eq}}(N)=\frac{1-\hat{\phi}(s)}{s\langle\tau\rangle} \hat{\phi}(s)^{N-1} \frac{1-\hat{\phi}(s)}{s}
\end{equation}
with $N>0$. Here, we focus on the short-time limit of Eq.~\eqref{equilibriumQTS}, which corresponds to $s\to\infty$.
\subsection{Generalized power-series expansion}
\begin{figure}[tp!]
 \centering
 \includegraphics[width=0.48\textwidth]{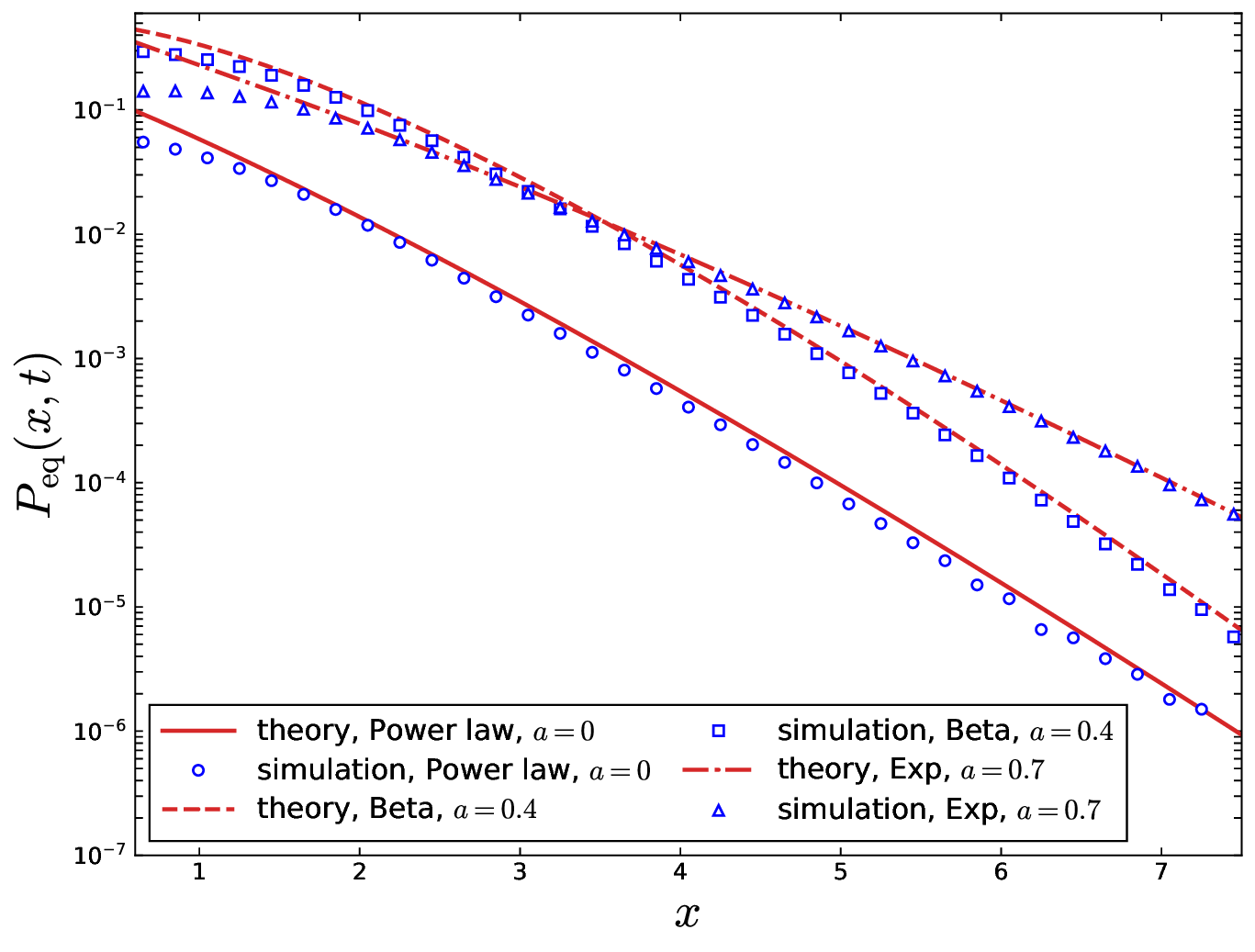}\vspace{-0.3cm}
 \caption{The far tails of the positional distribution under equilibrium conditions for different bias strengths are shown, with the bias values indicated in the legend. The lines represent the theoretical predictions from Eq.~\eqref{eqPxtAsy}, while the symbols correspond to the simulation results. For the displacements, we choose $\sigma=1$. Three waiting-time distributions are considered: a power-law distribution, $\phi(\tau)=\gamma/(1+\tau)^{1+\gamma}$ with $\gamma=3/2$ ($'\circ'$); a beta distribution, $\phi(\tau)=6\tau(1-\tau)$ ($'\Box'$); and an exponential distribution, $\phi(\tau)=\exp(-\tau)$ ($'\triangle'$). It can be seen that, for all these cases, the positional distributions exhibit exponential decay in the far-tail regime.
}\label{EqPxt}
\end{figure}

Our first example is the generalized power-series expansion given by  Eq.~\eqref{26Sec2101} but with a finite mean.
In the limit $s\to\infty$, applying the inverse Laplace transform with respect to $s$ to Eq.~\eqref{equilibriumQTS} leads to \cite{Wang2026Far}
\begin{equation}\label{QTNEQURIBRIUM}
\begin{aligned}
Q_t^{\mathrm{eq}}(N)\sim{}&\frac{t}{\langle\tau\rangle}\cdot
\frac{\left[\left(C_{\alpha}\Gamma(1+\alpha)\right)^{1/(1+\alpha)}t\right]^{(1+\alpha)(N-1)}}{\Gamma(2+(1+\alpha)(N-1))}
\\
&\times\exp\left[(N-1)^{1+\alpha-\beta}
\frac{C_{\beta}\Gamma(1+\beta)t^{\beta-\alpha}}{C_{\alpha}\Gamma(1+\alpha)(1+\alpha)^{\beta-\alpha}}\right],
\end{aligned}
\end{equation}
describing the far tail of the distribution of number of renewals. In the large $N$ limit, the far tail exhibits an approximately exponential decay. Utilizing Eqs.~\eqref{QTNEQURIBRIUM}  and \eqref{26subInter}, the far tails of the positional distribution follows
\[
P_{\rm eq}(x, t) \sim \int_{0}^{\infty} \exp \left(\begin{array}{c}
\ln \left(\frac{t}{\langle\tau\rangle}\right)+(1+\alpha)(N-1) \ln \left(\left(C_{\alpha} \Gamma(1+\alpha)\right)^{\frac{1}{1+\alpha}} t\right)\\
-\ln (\Gamma(2+(1+\alpha)(N-1)))-\frac{(x-a N)^{2}}{2 \sigma^{2} N} \\
-\frac{1}{2} \ln \left(2 \pi \sigma^{2} N\right) \\
+(N-1)^{1+\alpha-\beta} \frac{C_{\beta} \Gamma(1+\beta) t^{\beta-\alpha}}{C_{\alpha} \Gamma(1+\alpha)(1+\alpha)^{\beta-\alpha}}
\end{array}\right) dN,
\]
Similar to the above procedure, we use the saddle point method to calculate its asymptotic behavior. Taking the derivative with respect to $N$, we have
\begin{gather*}
\chi^{\prime}(N) \sim-\frac{1}{2 N}+(1+\alpha) \ln \left(\frac{\left[C_{\alpha} \Gamma(1+\alpha)\right]^{\frac{1}{1+\alpha}} t}{2+(1+\alpha)(N-1)}\right)+\frac{x^{2}}{2 \sigma^{2} N^{2}}-\frac{a^{2}}{2 \sigma^{2}} \\
+\frac{1+\alpha}{2[2+(1+\alpha)(N-1)]}+(1+\alpha-\beta)(N-1)^{\alpha-\beta} \frac{C_{\beta} \Gamma(1+\beta) t^{\beta-\alpha}}{C_{\alpha} \Gamma(1+\alpha)(1+\alpha)^{\beta-\alpha}}=0.
\end{gather*}
In the limit of large $N$, the asymptotic root of the above equation is
\begin{equation}\label{nXINGEQUAE}
N^{*} \approx \sqrt{\frac{\frac{x^{2}}{\sigma^{2}(1+\alpha)}}{W_0\left(\frac{x^{2} \exp \left(\frac{a^{2}}{\sigma^{2}(1+\alpha)}\right)}{\sigma^{2}(1+\alpha)\left(\frac{\left[C_{\alpha} \Gamma(1+\alpha)\right]^{\frac{1}{1+\alpha}} t}{1+\alpha}\right)^{2}}\right)}}.
\end{equation}
When the waiting times are drawn from Eq.~\eqref{26Sec2101}, the saddle point of the equilibrium CTRW coincides with that of the ordinary CTRW process given by Eq.~\eqref{NxingGeneralexpansion}. This indicates that the value of $N^*$ is insensitive to the statistics of the first waiting time.  Based on Eq.~\eqref{nXINGEQUAE}, we get the asymptotic solution of the positional distribution
\begin{equation}\label{eqPxtAsy}
\begin{aligned}
P_{\rm eq}(x,t)\sim{}&
\frac{
(1+\alpha)^{\frac{\alpha}{2}-\frac{3}{4}}
|x|^{\alpha-\frac{1}{2}}
W_0^{\frac{1-2\alpha}{4}}
\left(g_a\frac{x^2}{t^2}\right)
}{
\sqrt{2\pi}\,\langle\tau\rangle
C_\alpha\Gamma(\alpha+1)
\sigma^{\alpha+\frac{1}{2}}t^\alpha
\sqrt{
W_0\left(g_a\frac{x^2}{t^2}\right)+1
+R_1\left(\frac{t}{|x|}\right)^{\beta-\alpha}
W_0^{\frac{\beta-\alpha}{2}}
\left(g_a\frac{x^2}{t^2}\right)
}
}
\\[1mm]
&\times
\exp\Bigg[
-|x|G_a\left(\frac{x}{t}\right)
+\frac{a}{\sigma^2}x
+\sqrt{\alpha+1}\,R_2t
\left(\frac{t\sigma}{|x|}\right)^{\beta-\alpha-1}
W_0^{\frac{\beta-\alpha-1}{2}}
\left(g_a\frac{x^2}{t^2}\right)
\\
&\hspace{2.0cm}
-(\alpha+1)(\alpha-\beta+1)R_2
\left(\frac{t\sigma}{|x|}\right)^{\beta-\alpha}
W_0^{\frac{\beta-\alpha}{2}}
\left(g_a\frac{x^2}{t^2}\right)
\Bigg],
\end{aligned}
\end{equation}
where $g, R_2, W_0(t)$ are defined functions provided below Eq.~\eqref{26Sec2108}.
For large $x$, the leading term of Eq.~\eqref{eqPxtAsy} yields
\begin{equation}
P_{\rm eq}(x,t)\sim \exp\left[\frac{a}{\sigma^2}x-\frac{|x|}{\sigma}\sqrt{(1+\alpha)W_0(gl^2)}\right],
\end{equation}
which indicates an approximately exponential decay. See Fig.~\eqref{EqPxt} for a verification.

\subsubsection{Ratio of the tails}
\begin{figure}[tp!]
 \centering
\includegraphics[width=0.48\textwidth]{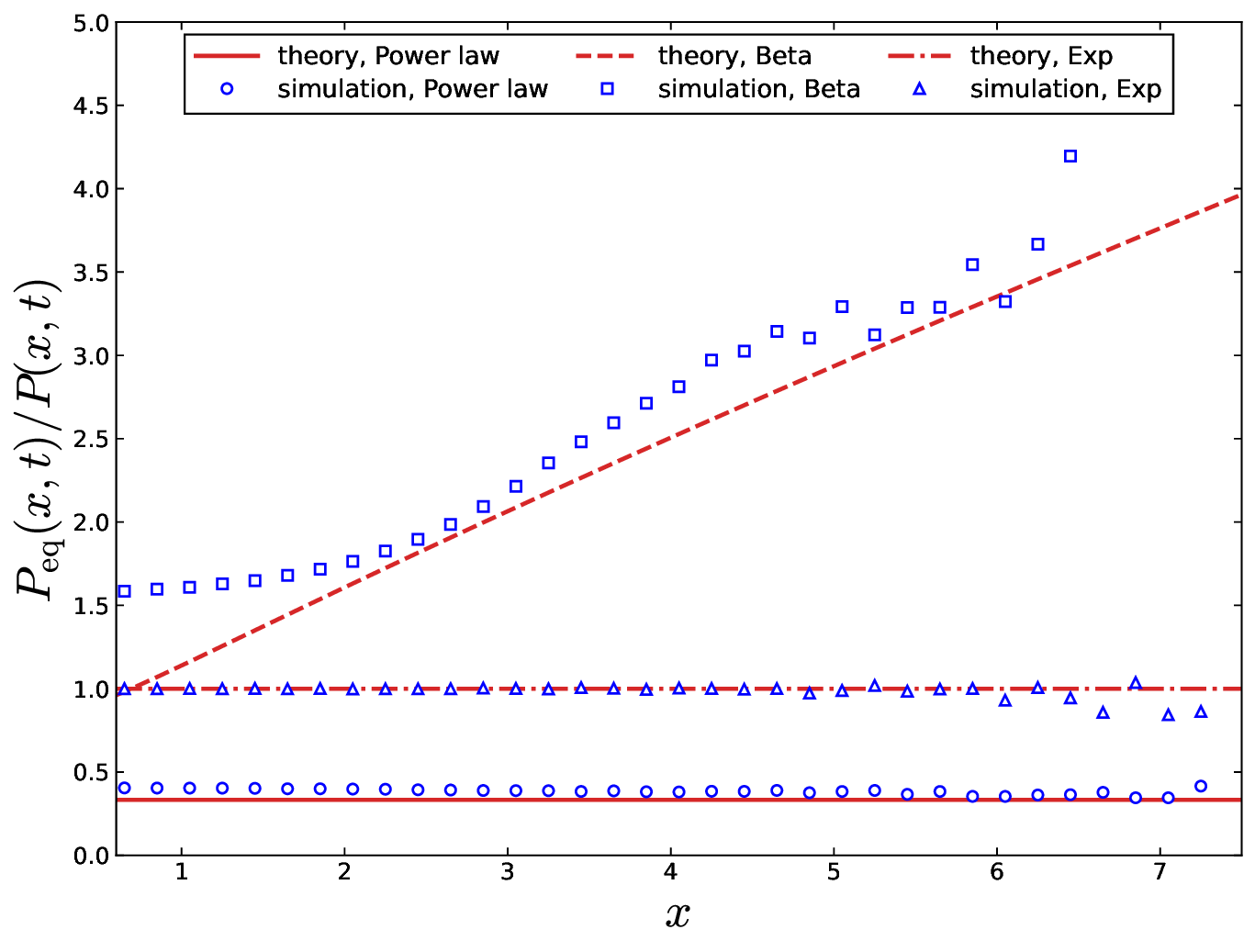}\vspace{-0.3cm}
 \caption{Ratio of the positional distributions under equilibrium and normal initial conditions. The lines represent the theoretical prediction given by Eq.~\eqref{eqPeqPAsy}, while the symbols denote simulation results. We set $a=0.1$ and $\sigma=1$. The statistics of waiting times are the same as the one given in Fig.~\ref{EqPxt}.
}\label{RatioEqVsNor}
\end{figure}

Based on Eq.~\eqref{eqPxtAsy}, we observe that the leading-order term is identical to that in Eq.~\eqref{AsyNorPpxt}. This naturally raises the question of whether the two initial conditions lead to the same tail statistics or whether the tails retain a dependence on the initial conditions. Below, we investigate this question in detail. Utilizing Eqs.~\eqref{26Sec2108} and \eqref{eqPxtAsy}, we find the ratio for the positional distribution
\begin{equation}
\begin{aligned}
\frac{P_{\rm eq}(x,t)}{P(x,t)}
\sim{}&
\frac{1}
{\langle\tau\rangle C_\alpha\Gamma(\alpha+1)}
\left[
\frac{
\sqrt{\alpha+1}\,|x/t|
}{
\sigma
\sqrt{
W_0\left(g_a\frac{x^2}{t^2}\right)
}
}
\right]^\alpha
\\
&\quad{}\times\exp\Bigg[
-
\frac{
C_\beta\Gamma(\beta+1)
(\alpha-\beta+1)
}{
C_\alpha\Gamma(\alpha+1)
(\alpha+1)^{\frac{\beta-\alpha}{2}}
}
\left(\frac{t\sigma}{|x|}\right)^{\beta-\alpha}
W_0^{\frac{\beta-\alpha}{2}}
\left(g_a\frac{x^2}{t^2}\right)
\Bigg].
\end{aligned}
\end{equation}
Using $W_0(y)\sim \ln y$ for $y\to+\infty$, we have
\begin{equation}
\begin{aligned}
\frac{P_{\rm eq}(x,t)}{P(x,t)}
\sim{}&
\frac{1}
{\langle\tau\rangle C_\alpha\Gamma(\alpha+1)}
\left[
\frac{
\sqrt{\alpha+1}\,|x/t|
}{
\sigma
\sqrt{
W_0\left(g_a\frac{x^2}{t^2}\right)
}
}
\right]^\alpha
\\
&\quad{}\times\exp\Bigg[
-
\frac{
C_\beta\Gamma(\beta+1)
(\alpha-\beta+1)
}{
C_\alpha\Gamma(\alpha+1)
(\alpha+1)^{\frac{\beta-\alpha}{2}}
}
\left(\frac{t\sigma}{|x|}\right)^{\beta-\alpha}
\ln^{\frac{\beta-\alpha}{2}}
\left(g_a\frac{x^2}{t^2}\right)
\Bigg].
\end{aligned}
\end{equation}
Taking the further limit $|x|/t\to\infty$ and the correction to the exponent vanishes asymptotically, leading to
\begin{equation}\label{eqPeqPAsy}
\frac{P_{\rm eq}(x,t)}{P(x,t)}
\sim
\frac{1}
{\langle\tau\rangle C_\alpha\Gamma(\alpha+1)}
\left[
\frac{
\sqrt{\alpha+1}\,|x/t|
}{
\sigma
\sqrt{
W_0\left(g_a\frac{x^2}{t^2}\right)
}
}
\right]^\alpha.
\end{equation}
When $\alpha>0$, the ratio increases with position, as illustrated for the beta waiting-time distribution in Fig.~\ref{RatioEqVsNor}. For large $x$, the asymptotic behavior of Eq.~\eqref{eqPeqPAsy} follows
\begin{equation}
\frac{P_{\rm eq}(x,t)}{P(x,t)}\propto
\left(\frac{|x|}{t\sqrt{W_0(g_ax^2/t^2)}}\right)^\alpha.
\end{equation}
In the special case $\alpha=0$, Eq.~\eqref{eqPeqPAsy} reduce to a constant
\begin{equation}\label{ratioAlpha0}
\frac{P_{\rm eq}(x,t)}{P(x,t)}\sim \frac{1}{\langle\tau\rangle C_0}.
\end{equation}
For this special case, the ratio of the positional distributions under equilibrium and ordinary conditions is determined by the mean waiting time and the leading coefficient in the Taylor expansion of the waiting-time distribution. When the waiting times are drawn from the exponential distribution, $\phi(\tau)=\exp(-\tau)$, the ratio is unity, as expected. In other words, the initial condition has no effect on the positional distribution, reflecting the absence of aging in the exponential waiting-time process. See Fig.~\ref{RatioEqVsNor} for simulations.

On the other hand, the ratio can be interpreted from a different perspective: we use $N^{*}$, which denotes the most likely number of steps required to reach the position $x$, to characterize the positional statistics. To this end, the positional distribution can be formally written by
\begin{equation}\label{skdns101}
P(x, t) \sim  Q_{t}\left(N^{*}\right) f\left(x \mid N^{*}\right)
\end{equation}
and
\begin{equation}\label{skdns102}
P_{e q}(x, t) \sim  Q_{t, e q}\left(N_{e q}^{*}\right) f\left(x \mid N_{e q}^{}\right).
\end{equation}
Here, $N^{*}$ and $N_{eq}^{*}$ denote the saddle points for the normal and equilibrium CTRW, respectively.
Thus, the ratio of the positional distribution obeys
\begin{equation*}
\frac{P_{e q}(x, t)}{P(x, t)} \sim \frac{Q_{t, e q}\left(N_{e q}^{*}\right) f\left(x \mid N_{e q}^{*}\right)}{Q_{t}\left(N^{*}\right) f\left(x \mid N^{*}\right)}.
\end{equation*}
Recall that $N^{*}$ and $N_{eq}^{*}$ coincide for the processes considered here. Then we have
\begin{equation}\label{RationNxing1001}
\frac{P_{e q}(x, t)}{P(x, t)} \sim \frac{Q_{t, e q}\left(N^{*}\right)}{Q_{t}\left(N^{*}\right)}\sim \frac{(1+\alpha)^\alpha}
{\langle\tau\rangle C_\alpha\Gamma(\alpha+1)}
\left(\frac{N^*}{t}\right)^\alpha.
\end{equation}
This indicates that the ratio of the positional distributions can be expressed as a function of $N^{*}/t$. It can be readily verified that Eq.~\eqref{RationNxing1001} is consistent with Eq.~\eqref{eqPeqPAsy}.

\subsubsection{Ratio of the positional distribution with and without bias}
Similar to the normal CTRW model, we further consider the positional distributions with and without bias. This ratio allows us to study the effect of the bias and examine whether it persists in the asymptotic tails. We therefore define the ratio $R_{\mathrm{eq}}(x,t)$ as
\begin{equation}
\mathcal{R}_{\mathrm{eq}}(x, t) = \frac{ P_{\mathrm{eq}}(x, t) \big|_{a \neq 0} }{ P_{\mathrm{eq}}(x, t) \big|_{a = 0}}.
\end{equation}
Based on Eq.~\eqref{eqPxtAsy}, we have
\begin{equation}\label{Rario261023}
\begin{aligned}
\mathcal{R}_{\rm eq}(x,t)
\sim{}&
\mathcal{R}(x,t)
\left[
\frac{
W_0\left(g_0\frac{x^2}{t^2}\right)
}{
W_0\left(g_a\frac{x^2}{t^2}\right)
}
\right]^{\alpha/2}
\\
&\quad{}\times\exp\Bigg[
-
\frac{
C_\beta\Gamma(\beta+1)(\alpha-\beta+1)
}{
C_\alpha\Gamma(\alpha+1)
(\alpha+1)^{\frac{\beta-\alpha}{2}}
}
\left(\frac{t\sigma}{|x|}\right)^{\beta-\alpha}
\\
&\qquad\quad{}\times\Bigg(
W_0^{\frac{\beta-\alpha}{2}}
\left(g_a\frac{x^2}{t^2}\right)
-
W_0^{\frac{\beta-\alpha}{2}}
\left(g_0\frac{x^2}{t^2}\right)
\Bigg)
\Bigg],
\end{aligned}
\end{equation}
where $\mathcal{R}(x,t)$ is defined in Eq.~\eqref{ratioOfPositionalBiased}.  Rewriting Eq.~\eqref{Rario261023}, we get a simple expression
\begin{equation}\label{RatioOfRatio}
\frac{\mathcal{R}_{\rm eq}(x,t)}{\mathcal{R}(x,t)}\sim \left[
\frac{
W_0\left(g_0\frac{x^2}{t^2}\right)
}{
W_0\left(g_a\frac{x^2}{t^2}\right)
}
\right]^{\alpha/2}\sim \left[
\log_{g_a \frac{x^2}{t^2}}\left(g_0\frac{x^2}{t^2}\right)
\right]^{\alpha/2}
\end{equation}
with $x\to \infty$. This relation is valid for arbitrary bias. Below, we consider a special case of Eq.~\eqref{RatioOfRatio}, corresponding to the weak-bias limit $a\to 0$. For small $a$, $g_a$ is an even function of $a$, and thus we can approximate
\[
W_0\left(g_a\frac{x^2}{t^2}\right)\simeq W_0\left(g_0\frac{x^2}{t^2}\right).
\]
Utilizing Eq.~\eqref{ratioOfPositionalBiased2} and the above equation, we obtain the leading-order behavior of the ratio as
\begin{equation}
\ln(\mathcal{R}(x,t)_{{\rm eq}})\sim\frac{a}{\sigma^2}x,
\end{equation}
which demonstrates a linear dependence of $\ln(\mathcal{R}(x,t)_{{\rm eq}})$ on the position $x$. The coefficient of this linear term, $a/\sigma^2$, characterizes the effect of the bias and is determined by the ratio between the mean displacement and the variance of the jump distribution.

\subsubsection{Rate function}
We now turn to the position rate function and the time rate function that characterizes rare-event statistics, which is derived in the limit \(x\to\infty\) and \(t\to\infty\), respectively.
From \eqref{eqPxtAsy}, we have
\begin{equation}\label{eqPxtAsy12}
\begin{aligned}
P_{\rm eq}(x,t)
\sim
\exp\Bigg[
&-|x|G_a\left(\frac{x}{t}\right)
+\frac{a}{\sigma^2}x+\sqrt{1+\alpha}\,R_2 t
\left(\frac{t\sigma}{|x|}\right)^{\beta-\alpha-1}
W_0^{\frac{\beta-\alpha-1}{2}}
\left(g_a\frac{x^2}{t^2}\right)
\Bigg].
\end{aligned}
\end{equation}
Rewriting the above equation, we obtain
\begin{equation}
\begin{aligned}
\ln P_{\rm eq}(x,t)
\sim{}&
-|x|G_a\left(\frac{x}{t}\right)
+\frac{a}{\sigma^2}x
+\sqrt{1+\alpha}\,R_2 t
\left(\frac{t\sigma}{|x|}\right)^{\beta-\alpha-1}
W_0^{\frac{\beta-\alpha-1}{2}}
\left(g_a\frac{x^2}{t^2}\right).
\end{aligned}
\end{equation}
Consequently, the position rate function $\mathcal{I}^{\rm{eq}}_x(l)$ follows
\begin{equation}\label{RatxEqPosition}
\begin{aligned}
\frac{\ln P_{\rm eq}(x,t)}{-|x|}
\sim \mathcal{I}^{\rm{eq}}_x(\frac{x}{t})
\end{aligned}
\end{equation}
with
\begin{equation}
\begin{aligned}
\mathcal{I}^{\rm{eq}}_x(l)
=G_a(l)
-\frac{a}{\sigma^2}\operatorname{sign}(l)
-\sqrt{1+\alpha}\,R_2
\frac{
\sigma^{\beta-\alpha-1}
}{
|l|^{\beta-\alpha}
}
W_0^{\frac{\beta-\alpha-1}{2}}
\left(g_a l^2\right).
\end{aligned}
\end{equation}
Similarly, the time rate function is given by
\begin{equation}\label{RatxEqTime}
\begin{aligned}
\lim_{t\to\infty}\frac{\ln P_{{\rm eq}}(x,t)}{-t}&=\mathcal{I}^{\rm{eq}}_t(l) \\
&=|l|G_a(l)
-\frac{a}{\sigma^2}l \\
&\quad{}-
\frac{
C_\beta\Gamma(\beta+1)
\sigma^{\beta-\alpha-1}
}{
C_\alpha\Gamma(\alpha+1)
(1+\alpha)^{\frac{\beta-\alpha+1}{2}}
}
|l|^{1-\beta+\alpha}
W_0^{\frac{\beta-\alpha-1}{2}}
\left(g_a l^2\right).
\end{aligned}
\end{equation}
Based on Eqs.~\eqref{RatxEqPosition} and \eqref{RatxEqTime}, we find that the rate functions are identical to those of the ordinary CTRW. This indicates that the initial conditions considered in this manuscript do not affect the rate functions.

\subsection{Waiting-time distribution with a cutoff}
We further consider the case when distributions of waiting times have a cutoff at $\tau=\tau_0$, whose expansion form is the same as Eq.~\eqref{cutoff101} but with a finite mean. At the short time limit, the law of the number of renewals follows \cite{Wang2026Statistics}
\begin{equation}\label{sdesd}
\begin{aligned}
Q_{t}^{\rm eq}(N)\sim{}& \frac{t-\tau_0(N-1)}{\langle\tau\rangle} \\
&\quad{}\times\frac{[(C_\alpha \Gamma(1+\alpha))^{\frac{1}{1+\alpha}}(t-\tau_0(N-1))]^{(1+\alpha)(N-1)}}{\Gamma(2+(1+\alpha)(N-1))} \\
&\quad{}\times\exp\left(\frac{C_\beta\Gamma(1+\beta)(t-\tau_0(N-1))^{\beta-\alpha}}{C_\alpha \Gamma(1+\alpha)(1+\alpha)^{\beta-\alpha}/(N-1)^{1+\alpha-\beta}}\right)
\end{aligned}
\end{equation}
with $N<1+t/\tau_0$.
However, the far tail of the number of renewals is not exponential and exhibits a cutoff. Then the far tail of the distribution of the position is
\begin{equation}\label{EquriPPPPxcutoff}
P_{{\rm eq}}(x,t)\sim \sum_{N=1}^{\lfloor t/\tau_0+1 \rfloor }Q_{t}^{\rm eq}(N)f(x|N).
\end{equation}
Although the waiting-time distribution in Eq.~\eqref{26Sec2101} is modified by a simple shift $\tau_0$, the saddle-point equation remains difficult to solve directly because of the nontrivial coupling between $x$ and $N$. In \ref{app:explicit-saddle}, we address this difficulty by employing a perturbative approximation \cite{bender1999advanced}, also known as the bookkeeping parameter method, to derive the following asymptotic expression for the far tails of the positional distribution.
\begin{equation}\label{eq:perturbative-propagator}
P_{\mathrm{app}}(x,t)=
\sqrt{\frac{2\pi}{\Phi''(N_{\mathrm{app}})}}\,
\exp[-\Phi(N_{\mathrm{app}})],
\end{equation}
where $\Phi(\cdot)$ and $N_{\mathrm{app}}$ are defined in Eqs.~\eqref{CufooffPhisn} and \eqref{equssjdssd} of \ref{app:explicit-saddle}, respectively. See Figs.~\ref{fig:exponential-propagator} and
\ref{fig:dagum-propagator} in Appendix for the results corresponding to the exponential and Dagum distributions, respectively.

Compared with Eq.~\eqref{QTNEQURIBRIUM}, the number of renewals cannot tend to infinity  because of the lower cutoff $\tau_0$ imposed on the waiting times. This raises the question of the role played by the maximum possible number of renewals, defined as $N_{\max}=\lfloor t/\tau_0+1\rfloor$. Motivated by the approximation in Eq.~\eqref{skdns102}, we find that the far tails of the positional distribution near its edge follow
\begin{equation}\label{eqssBigNum}
P_{\mathrm{eq}}(x,t)
\sim
\frac{Q_t^{\mathrm{eq}}(N_{\max})}{\sqrt{2\pi\sigma^2N_{\max}}}
\exp\!\left[-\frac{(x-aN_{\max})^2}{2\sigma^2N_{\max}}\right],
\qquad |x|\to \infty.
\end{equation}
It can be seen that, instead of using the finite sum in Eq.~\eqref{EquriPPPPxcutoff} to compute the large-$x$ behavior, we can use a fixed number of renewals.
The Gaussian far-tail behavior describes particles reaching extremely large positions. When the waiting-time distribution has a cutoff, the number of renewals is also bounded. In contrast, the particle position has no finite upper bound, even mathematically, and can therefore extend to infinity. It indicates that this behavior differs significantly from that of the first family of waiting-time distributions given in Eq.~\eqref{26Sec2101}. As illustrated in the plots of Figs.~\ref{fig:exponential-propagator} and \ref{fig:dagum-propagator}, the aforementioned far tails are illustrated.

\section{Conclusion}

In this manuscript, we investigate the far-tail behavior of CTRW propagators in the short-observation-time regime using the saddle-point method. We derive the ordinary and equilibrium propagators for different waiting-time distributions by expressing the positional propagator as a sum over the number of renewals. The asymptotic behavior of the renewals is then used to identify the dominant saddle point, which determines the most likely number of renewals required to reach a given position. This approach yields both the decay exponent and its associated coefficient, establishing a direct connection between the short-time behavior of the waiting-time density and the spatial far tails of the propagator.

For waiting-time densities that admit a generalized power-series expansion near the origin, we extend the analysis to non-integer powers, including densities that diverge at the origin. In these cases, the saddle point can be expressed  in terms of the Lambert $W$ function, and the propagator exhibits nearly exponential decay with logarithmic corrections. In the weak-bias regime, the logarithm of the ratio of the biased to unbiased positional distributions, $\ln\mathcal{R}(x,t)$, is linear in position in the far tails, with a slope of $a/\sigma^2$. Remarkably, this relationship is independent of both the waiting-time statistics and the observation time $t$, making it potentially useful for real-world experiments. Furthermore, the far-tail behavior is insensitive to the asymptotic form of the waiting-time distribution at large waiting times. Thus, the rare fluctuations considered here differ from typical fluctuations, which are generally governed by the mean or variance of the waiting-time distribution. In the presence of a lower cutoff in the waiting-time density, the far tails of the positional distribution are strongly influenced by the maximum possible number of renewals, leading to an unexpected Gaussian decay.

The exponential decay of the propagator changes when extremely short waiting times are more strongly suppressed. In particular, for the one-sided L{\'e}vy stable waiting-time distributions, the propagator exhibits compressed-exponential decay. For a stability index $0<\alpha<1$, the leading decay exponent scales with the displacement as $|x|^{2/(2-\alpha)}$, yielding a power of $4/3$ when $\alpha=1/2$. Thus, even for the same Gaussian jump distribution, different behaviors of the waiting-time density near the origin can give rise to distinct spatial tail asymptotics. These differences reflect the probability of completing a large number of renewals within a short observation time.

We further investigate the CTRW model with equilibrium initial conditions. Although the saddle points are identical for the equilibrium and ordinary processes, their positional distributions exhibit different far-tail behaviors. More specifically, the ratio of the equilibrium and ordinary positional-distribution tails coincides with the corresponding ratio for the distributions of the number of renewals (see Eq.~\eqref{RationNxing1001}). When the waiting-time distribution has a cutoff near its minimum value, the far tails of the positional distribution exhibit Gaussian decay, in contrast to the renewal-count distribution, which has a cutoff in its far tail. Finally, we employ a perturbative approximation to investigate the saddle points in the presence of a cutoff in the waiting-time distribution. In this manuscript, we develop a saddle-point method to estimate the far tails of the positional distribution, which is applicable to a broad class of waiting-time distributions. Our approach provides a better understanding of rare events in the far-tail regime.

Bias breaks the symmetry between positive and negative displacements and modifies both the dominant renewal count and the corresponding far-tail probabilities. By comparing biased and unbiased propagators, we quantified the influence of bias on the far tail. In a weak-bias expansion, the first-order term in the logarithm of their ratio is proportional to the displacement. Besides, the results for different waiting-time distributions and initial conditions provide a common framework for understanding how renewal statistics and spatial bias determine the propagator at short observation times.

In this manuscript, we focus on the process with ordinary and equilibrium conditions. A natural extension is to consider process with a finite aging time, allowing the system to evolve for a finite preparation time before observation begins, and to investigate how the propagator tail depends on this preparation time. For waiting-time distributions with a finite mean, this setting can connect ordinary and equilibrium initial conditions. For distributions with a divergent mean, a finite-age process remains well defined, allowing the dependence of the far tail on the preparation history to be explored.

\appendix
\renewcommand{\theequation}{\Alph{section}\arabic{equation}}

\section{Perturbative approximation to the saddle point for the cutoff case}
\label{app:explicit-saddle}
Here, we provide an approximate form of the saddle-point starting from Eq.~\eqref{EquriPPPPxcutoff}. Throughout this appendix, $Q_t^{\mathrm{eq}}(N)$ is evaluated using the asymptotic expression in Eq.~\eqref{sdesd}, and the jump lengths are assumed to be Gaussian distributed with mean $a$ and variance $\sigma^2$. Similar to Eq.~\eqref{26Sec2103} for $\chi(N)$, we obtain
\begin{equation}\label{CufooffPhisn}
\Phi(N)=-\ln Q_t^{\mathrm{eq}}(N)
+\frac12\ln(2\pi\sigma^2N)
+\frac{(x-aN)^2}{2\sigma^2N},
\qquad \Phi'(N^*)=0.
\end{equation}
For brevity, we define
\[
\gamma=1+\alpha,\qquad A=C_\alpha\Gamma(1+\alpha),
\qquad u(N)=t-\tau_0(N-1),
\]
and denote the \(\beta\)-dependent exponent in Eq.~\eqref{sdesd} by
\[
\chi(N)=
\frac{C_\beta\Gamma(1+\beta)}{A\gamma^{\beta-\alpha}}
u(N)^{\beta-\alpha}(N-1)^{1+\alpha-\beta}.
\]
For fixed \(\gamma>0\) and \(\gamma N\gg1\),
\[
\gamma\Psi\!\left(2+\gamma(N-1)\right)+\frac1{2N}
\sim \gamma\ln(\gamma N)+\frac{2-\gamma}{N},
\]
where \(\Psi\) is the digamma function used in the main text. Consequently,
\begin{equation}\label{cutoffMain101}
\begin{aligned}
\Phi'(N)\sim{}&
\gamma\ln\frac{\gamma N}{A^{1/\gamma}u(N)}
+\frac{a^2}{2\sigma^2}
+\frac{\gamma\tau_0N}{u(N)}
-\frac{x^2}{2\sigma^2N^2}+\frac{2-\gamma}{N}
+\frac{(1-\gamma)\tau_0}{u(N)}
-\chi'(N)=0.
\end{aligned}
\end{equation}
In general, it is difficult to obtain the asymptotic solution of Eq.~\eqref{cutoffMain101} with all parameters included. Based on Eq.~\eqref{cutoffMain101}, we
construct  the following  reference equation
\begin{equation}\label{sassw121}
\frac{x^2}{2\sigma^2N^{\,2}}
=\gamma\ln\frac{\gamma N}
{A^{1/\gamma}(t+\tau_0)}
+\frac{a^2}{2\sigma^2}.
\end{equation}
Using the properties of the above equation, we obtain its solution as
\[
N_{\mathrm L}=\frac{|x|}{\sigma\sqrt{\gamma W}},
\qquad
W=W_0\!\left[
\frac{\gamma x^2}{\sigma^2(t+\tau_0)^2A^{2/\gamma}}
\exp\!\left(\frac{a^2}{\gamma\sigma^2}\right)
\right],
\]
where \(W_0\) denotes the principal branch of the Lambert function. The right-hand side of the reference equation equals \(\gamma W/2\). Below, we consider the terms discarded in Eq.~\eqref{sassw121} compared with Eq.~\eqref{cutoffMain101}, and define
\begin{equation}\label{eqwaA3}
H(N)=\gamma\left[\frac W2+\frac{\tau_0N}{u(N)}\right]
-\frac{x^2}{2\sigma^2N^2},
\qquad R(N)=\Phi'(N)-H(N).
\end{equation}
The difference between the actual logarithmic part and its reference value, the finite-\(N\) corrections, and the \(\beta\)-dependent term are all included in \(R(N)\). Equation~\eqref{eqwaA3} gives
\begin{equation}\label{eqsjsA4}
\begin{aligned}
R(N)\sim{}&
\gamma\ln\frac{\gamma N}{A^{1/\gamma}u(N)}
+\frac{a^2}{2\sigma^2}-\frac{\gamma W}{2}+\frac{2-\gamma}{N}
+\frac{(1-\gamma)\tau_0}{u(N)}
-\chi'(N).
\end{aligned}
\end{equation}
Equation~\eqref{eqwaA3} gives the exact remainder, whereas Eq.~\eqref{eqsjsA4} provides its large-\(N\) approximation. Either expression can be used in the perturbative formulas below, depending on the desired level of accuracy.

The zeroth-order root is determined by \(H(N_0)=0\), or equivalently,
\begin{equation}\label{eqs929291}
\tau_0(2-W)N_0^3+W(t+\tau_0)N_0^2
+\frac{\tau_0x^2}{\gamma\sigma^2}N_0
-\frac{(t+\tau_0)x^2}{\gamma\sigma^2}=0.
\end{equation}
Recall that, since we are interested in the large-\(x\) limit, we assume \(W>2\) throughout this analysis.
Since \(H'(N)>0\), this equation has a unique root in \(0<N_0<(t+\tau_0)/\tau_0\) for \(x\ne0\) and \(\tau_0>0\). Let
\[
\ell=\frac{\tau_0^2x^2}{2\gamma\sigma^2(t+\tau_0)^2},
\qquad b=\frac W2-\ell.
\]
Then
\[
N_0=\frac{t+\tau_0}{\tau_0}\frac{y}{1+y},
\qquad y^3+by^2-2\ell y-\ell=0.
\]
In the range \(W>2\) considered here, the cubic is positive at \(y=-1\) and negative at \(y=0\). Together with its behavior at infinity, these signs imply three distinct real roots, one in each of \((-\infty,-1)\), \((-1,0)\), and \((0,\infty)\). The unique positive root is
\[
y=
\frac{
2\sqrt{b^2+6\ell}
\cos\!\left[
\frac13\arccos\!\left(
\frac{27\ell-18b\ell-2b^3}
{2(b^2+6\ell)^{3/2}}
\right)\right]-b
}{3},
\]
where the function \(\arccos\) denotes its principal value. Thus, \(N_0\) is a root of the cubic cutoff equation, whereas \(N_{\mathrm L}\) serves only as the Lambert reference value used to construct this root.

Introduce a bookkeeping parameter \(\varepsilon\), and consider
\[
H(N)+\varepsilon R(N)=0,
\qquad
N=N_0+\varepsilon N_1+\varepsilon^2N_2+\cdots.
\]
Substitution and comparison of equal powers of \(\varepsilon\) give
\[
N_1=-\frac{R(N_0)}{H'(N_0)},
\qquad
N_2=
\frac{R(N_0)R'(N_0)}{[H'(N_0)]^2}
-\frac{H''(N_0)[R(N_0)]^2}{2[H'(N_0)]^3}.
\]
Setting \(\varepsilon=1\) yields the second-order approximation
\begin{equation}\label{equssjdssd}
\boxed{
N^*\simeq N_{\mathrm{app}}
=N_0-\frac{R(N_0)}{H'(N_0)}
+\frac{R(N_0)R'(N_0)}{[H'(N_0)]^2}
-\frac{H''(N_0)[R(N_0)]^2}{2[H'(N_0)]^3}.
}
\end{equation}
All corrections are evaluated at the same zeroth-order root \(N_0\); the first-order approximation is \(N_0+N_1\). The required derivatives are
\[
H'(N)=
\frac{\gamma\tau_0(t+\tau_0)}{u(N)^2}
+\frac{x^2}{\sigma^2N^3},
\qquad
H''(N)=
\frac{2\gamma\tau_0^2(t+\tau_0)}{u(N)^3}
-\frac{3x^2}{\sigma^2N^4},
\]
and
\begin{equation}\label{equssaws}
R'(N)\sim
\frac{\gamma}{N}+\frac{\gamma\tau_0}{u(N)}
-\frac{2-\gamma}{N^2}
+\frac{(1-\gamma)\tau_0^2}{u(N)^2}
-\chi''(N).
\end{equation}
The first- and second-order derivatives of the \(\beta\)-dependent term are given by
\[
\chi'(N)=
\frac{C_\beta\Gamma(1+\beta)}{A\gamma^{\beta-\alpha}}
u(N)^{\beta-\alpha-1}(N-1)^{\alpha-\beta}
\big[(1+\alpha-\beta)t-\tau_0(N-1)\big],
\]
and
\[
\chi''(N)=
\frac{C_\beta\Gamma(1+\beta)(\beta-\alpha)(\beta-\alpha-1)t^2}
{A\gamma^{\beta-\alpha}}
u(N)^{\beta-\alpha-2}(N-1)^{\alpha-\beta-1}.
\]
Dropping the remainders in Eqs.~\eqref{eqsjsA4} and \eqref{equssaws} gives formulas without digamma functions. Alternatively, using the full definition in Eq.~\eqref{eqwaA3}, with \(R'(N)=\Phi''(N)-H'(N)\), avoids this additional large-\(N\) truncation. The perturbative expansion requires \(N_0>1\), corrections confined to a local neighborhood of \(N_0\), and a corrected root within the physical interval with positive curvature. The bookkeeping parameter \(\varepsilon\) does not by itself guarantee convergence at \(\varepsilon=1\). The sizes of successive corrections, together with \(R'/H'\) and \(H''N_1/H'\), provide checks on the local expansion. If Eqs.~\eqref{eqsjsA4} and \eqref{equssaws} are used, one also requires \(\gamma N_0\gg1\). Relative accuracy of the propagator further requires the remaining root error to be small compared with the width \(w=[\Phi''(N^*)]^{-1/2}\), with controlled variation of the curvature. Applying the continuum saddle-point formula to the integer sum still requires the width, endpoint-distance, and higher-order conditions discussed in the main text.

\begin{figure}[ht]
\centering
\begin{minipage}{0.49\textwidth}
\centering
\includegraphics[width=\linewidth]{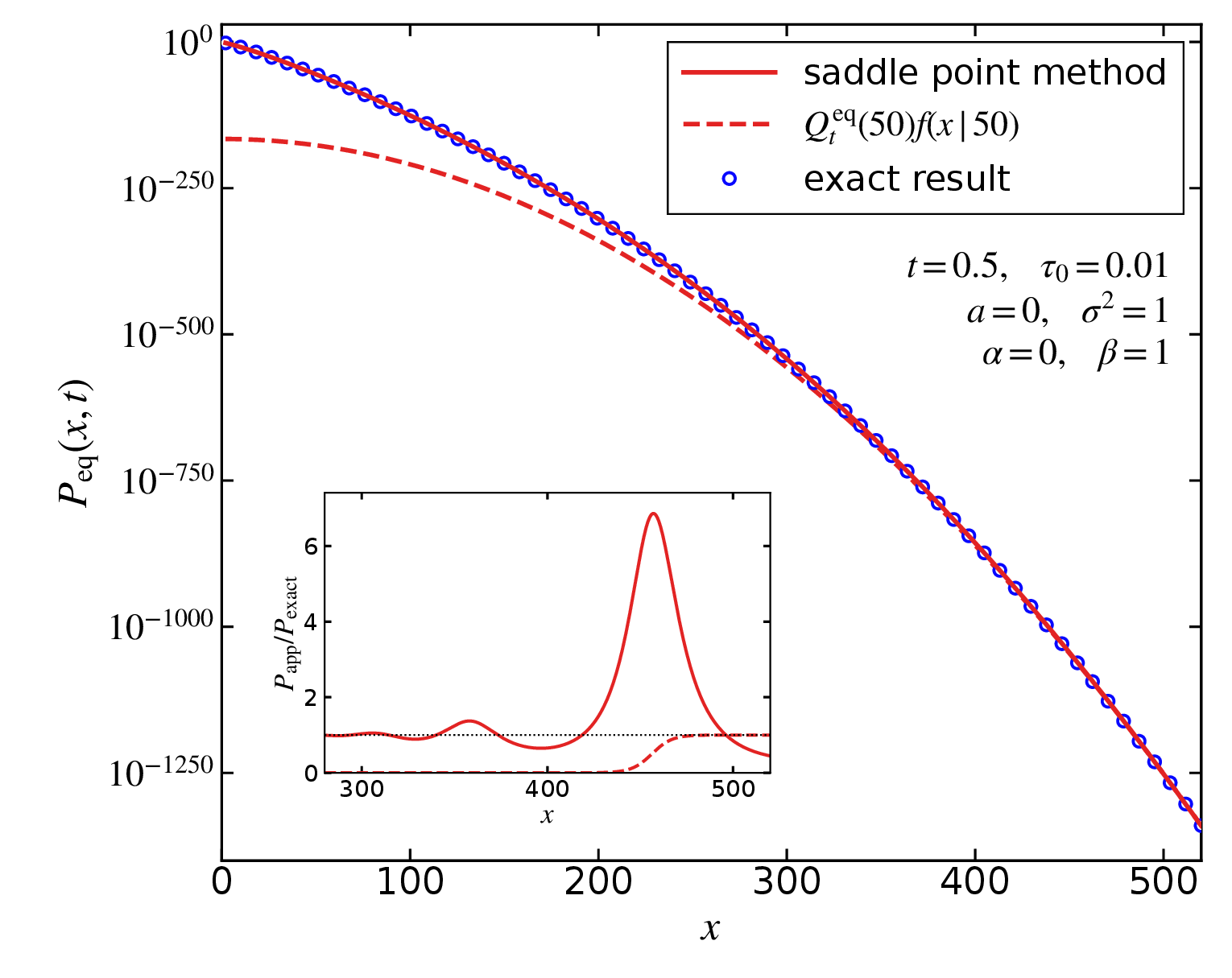}
\par\small (a) \(a=0\)
\end{minipage}\hfill
\begin{minipage}{0.49\textwidth}
\centering
\includegraphics[width=\linewidth]{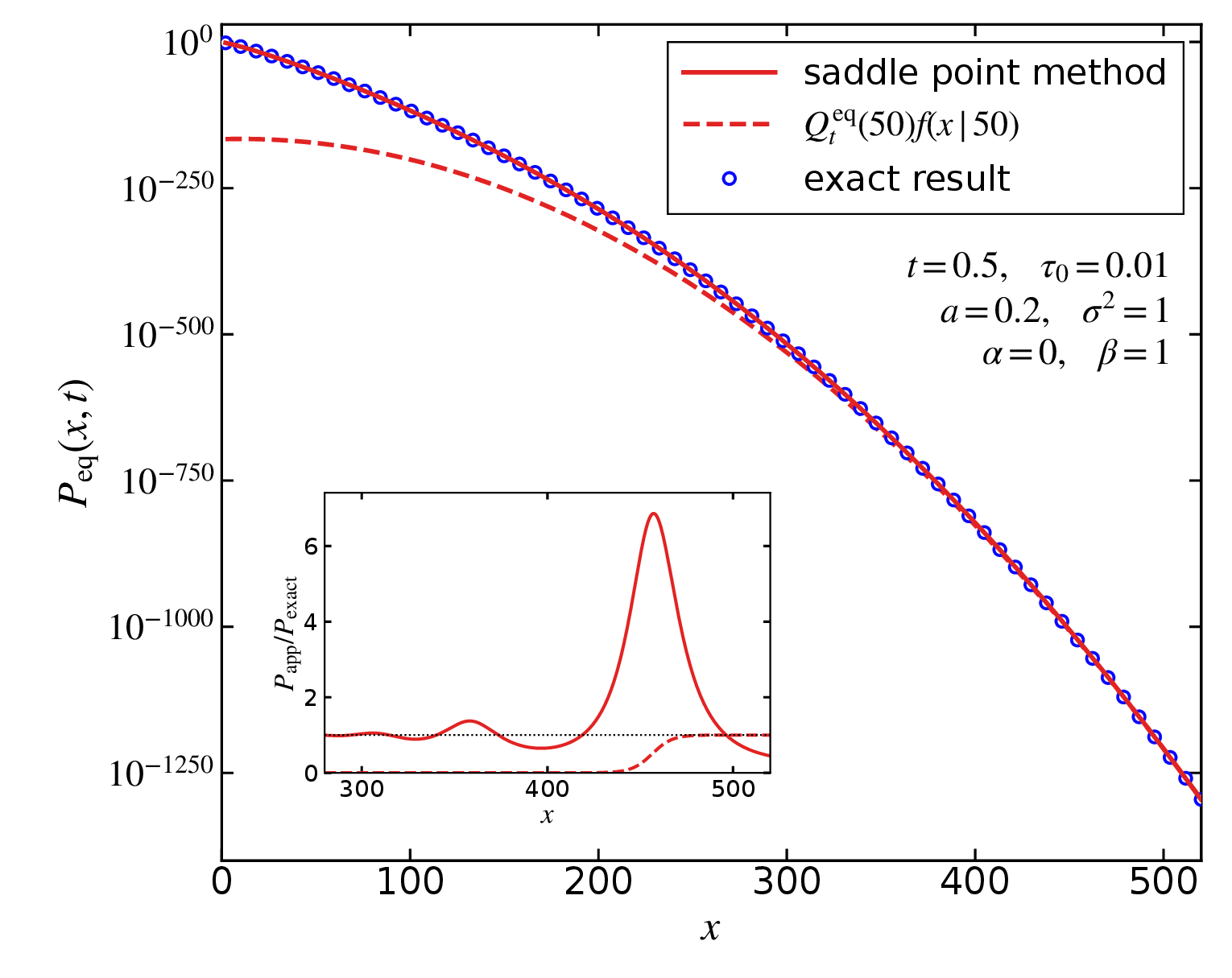}
\par\small (b) \(a=0.2\)
\end{minipage}
\caption{Equilibrium propagator for the shifted exponential waiting-time density, with \(t=0.5\), \(\tau_0=0.01\), and \(\sigma^2=1\). Blue open circles show the full numerical sum, red solid curves the numerical saddle-point approximation, and red dashed curves the single largest-count term. The insets show the ratios of the two approximations to the full sum; the dotted horizontal line marks unity. The broad logarithmic density range in the main panels conceals relative deviations that remain visible in the insets. These are deterministic numerical results, not Monte Carlo samples.}
\label{fig:exponential-propagator}
\end{figure}

\begin{figure}[ht]
\centering
\begin{minipage}{0.49\textwidth}
\centering
\includegraphics[width=\linewidth]{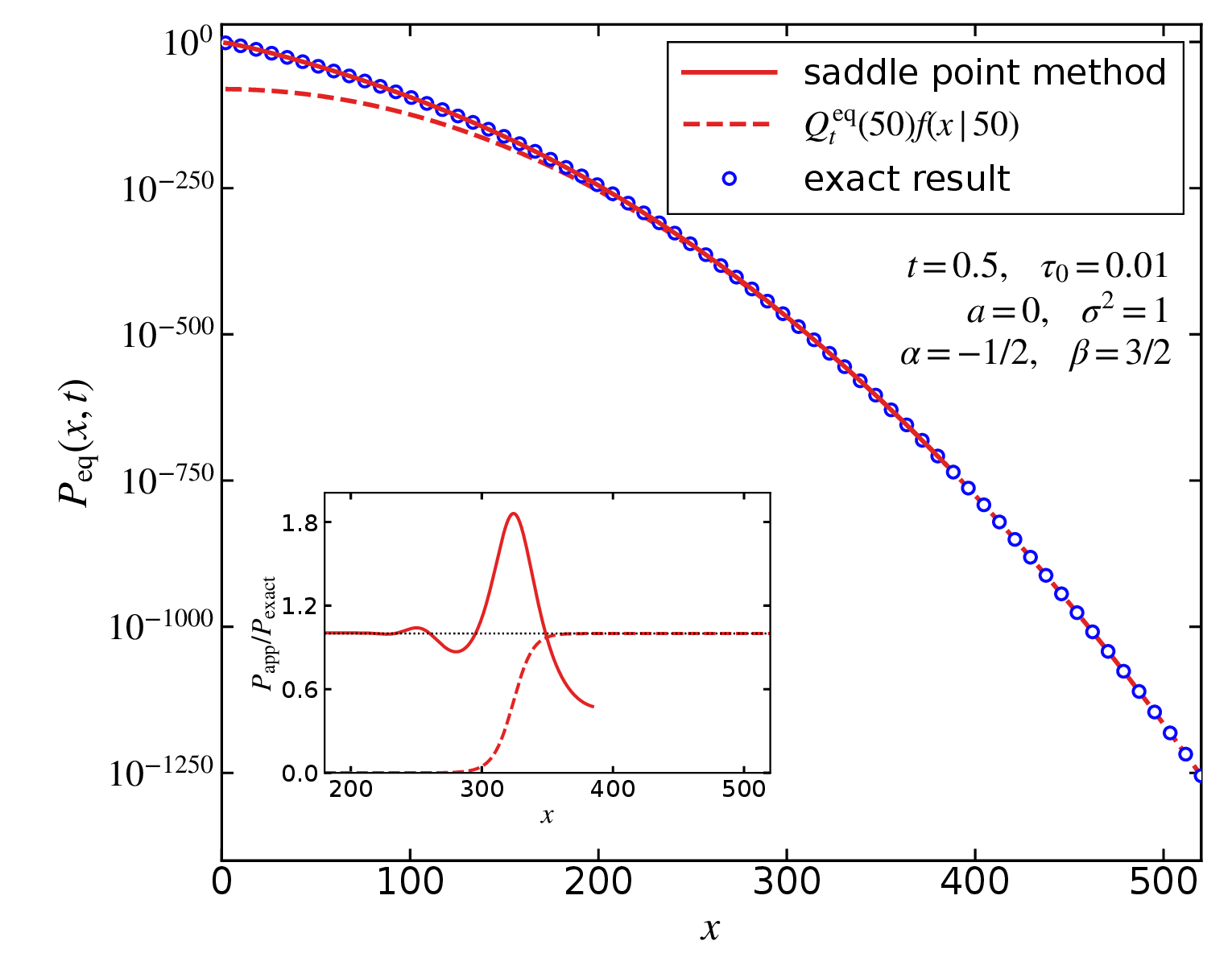}
\par\small (a) \(a=0\)
\end{minipage}\hfill
\begin{minipage}{0.49\textwidth}
\centering
\includegraphics[width=\linewidth]{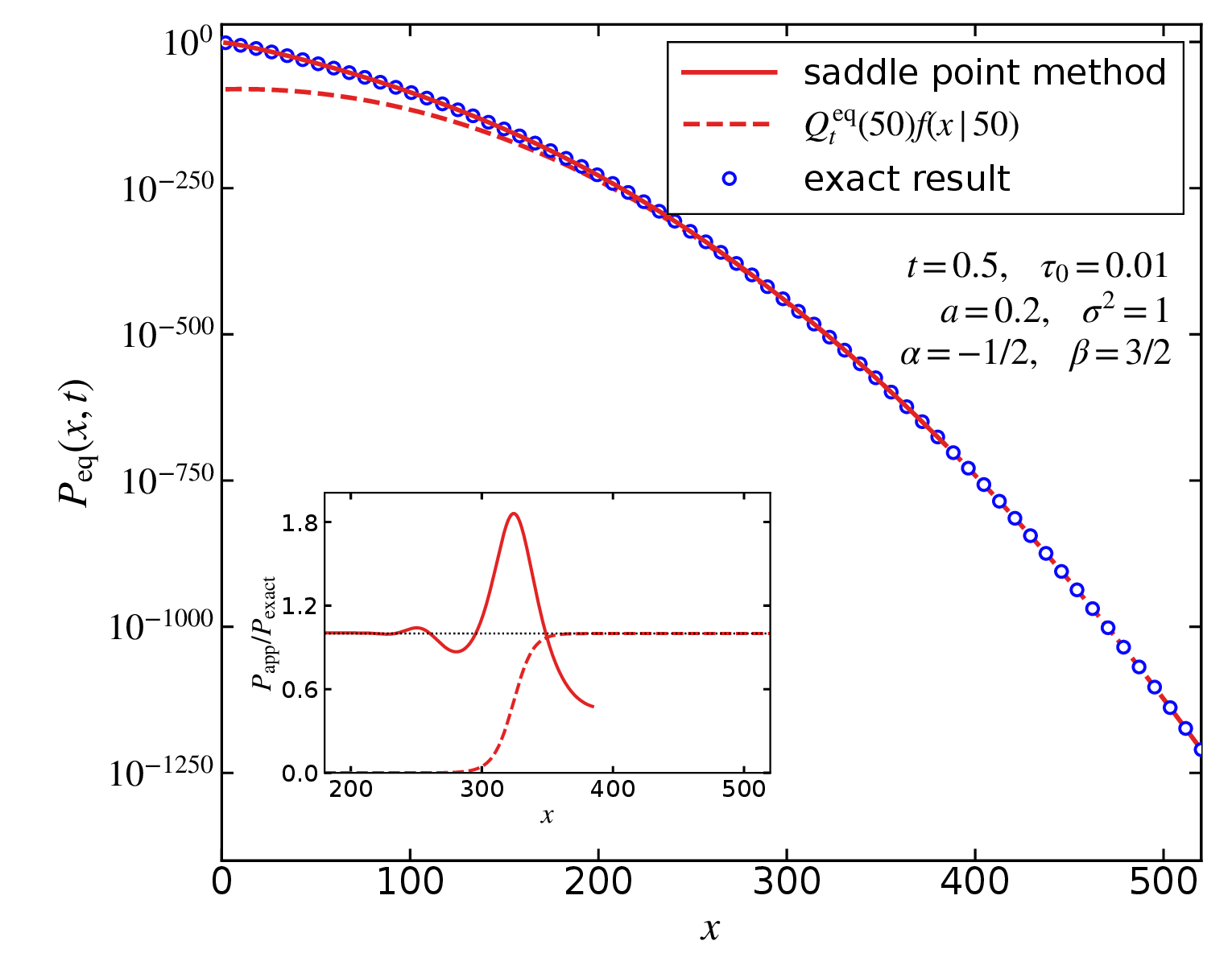}
\par\small (b) \(a=0.2\)
\end{minipage}
\caption{Equilibrium propagator for the shifted Dagum waiting-time density, with the same observation time, cutoff, jump variance, and curve definitions as in Fig.~\ref{fig:exponential-propagator}. The dashed curves are the single exact largest-count contribution, not the saddle-point formula evaluated at \(N=50\). The solid curves stop when the continuous stationary point reaches the upper end of the integer summation interval: \(x\simeq384.46\) for \(a=0\) and \(x\simeq384.59\) for \(a=0.2\). The largest-count term approaches the full sum in the ultimate far tail.}
\label{fig:dagum-propagator}
\end{figure}

For a numerical check, take
\[
\begin{gathered}
t=0.1,\qquad \tau_0=10^{-4},\qquad
\alpha=-\tfrac12,\qquad \beta=\tfrac12,\\
C_\alpha=\frac1{\sqrt\pi},\qquad C_\beta=-C_\alpha,
\qquad a=0,\qquad\sigma=1.
\end{gathered}
\]
The results below use the large-\(N\) expressions in Eqs.~\eqref{eqsjsA4} and \eqref{equssaws} to construct the approximate roots, which are then substituted into Eq.~\eqref{eq:perturbative-propagator}. The relative error is defined as \(100(P_{\mathrm{app}}/P_{\mathrm{sp,num}}-1)\%\), where \(P_{\mathrm{sp,num}}\) is obtained from the numerical saddle point of the same full action based on Eq.~\eqref{sdesd}.
\begin{center}
\begin{tabular}{rcc}
\hline
\(x\) & First-order relative error & Second-order relative error\\
\hline
\(1000\) & \(-1.5336\%\) & \(+0.02615\%\)\\
\(2000\) & \(-9.7471\%\) & \(-0.19679\%\)\\
\(3000\) & \(-3.4914\%\) & \(+0.001895\%\)\\
\(6000\) & \(-0.61469\%\) & \(-0.000826\%\)\\
\hline
\end{tabular}
\end{center}
Among the scanned points satisfying \(w\ge1\) and \(N^*\ge20\), the maximum absolute relative errors of the first- and second-order propagator approximations were approximately \(10.8\%\) and \(0.21\%\), respectively. This comparison tests the accuracy of replacing the numerical saddle point by an explicit approximate root. It does not include the error in Eq.~\eqref{sdesd} itself or the error arising from replacing the integer sum by a continuous integral.

\section*{References}
\bibliographystyle{tails-iop}
\bibliography{wenxian}

\providecommand{\newblock}{}
\begin{thebibliography}{10}
\expandafter\ifx\csname url\endcsname\relax
  \def\url#1{{\tt #1}}\fi
\expandafter\ifx\csname urlprefix\endcsname\relax\def\urlprefix{URL }\fi
\providecommand{\eprint}[2][]{\url{#2}}

\bibitem{Krapf2016Strange}
Krapf D, Campagnola G, Nepal K and Peersen O~B 2016 Strange kinetics of
  bulk-mediated diffusion on lipid bilayers {\em Phys. Chem. Chem. Phys.\/}
  {\bf 18}(18) 12633--12641
  \urlprefix\url{http://dx.doi.org/10.1039/C6CP00937A}

\bibitem{Metzler2000random}
Metzler R and Klafter J 2000 The random walk's guide to anomalous diffusion: a
  fractional dynamics approach {\em Phys. Rep.\/} {\bf 339} 1--77 ISSN
  0370-1573 \urlprefix\url{https://doi.org/10.1016/S0370-1573(00)00070-3}

\bibitem{Li2026Biased}
Li G, Cherstvy A~G, Guo W, Zhang H, Liu T, Pu W and Metzler R 2026 Biased
  continuous-time random walks with asymmetric waiting-time distributions as a
  model of anomalous diffusion in the presence of drift {\em Phys. Rev. E\/}
  {\bf 113}(6) 064116
  \urlprefix\url{https://link.aps.org/doi/10.1103/6c55-c9tp}

\bibitem{Hou2018Biased}
Hou R, Cherstvy A~G, Metzler R and Akimoto T 2018 Biased continuous-time random
  walks for ordinary and equilibrium cases: facilitation of diffusion{,}
  ergodicity breaking and ageing {\em Phys. Chem. Chem. Phys.\/} {\bf 20}(32)
  20827--20848 \urlprefix\url{http://dx.doi.org/10.1039/C8CP01863D}

\bibitem{Afek2003Colloquium}
Afek G, Davidson N, Kessler D~A and Barkai E 2023 Colloquium: Anomalous
  statistics of laser-cooled atoms in dissipative optical lattices {\em Rev.
  Mod. Phys.\/} {\bf 95}(3) 031003
  \urlprefix\url{https://link.aps.org/doi/10.1103/RevModPhys.95.031003}

\bibitem{Aquino2017Chemical}
Aquino T and Dentz M 2017 Chemical continuous time random walks {\em Phys. Rev.
  Lett.\/} {\bf 119}(23) 230601
  \urlprefix\url{https://link.aps.org/doi/10.1103/PhysRevLett.119.230601}

\bibitem{Kelly2019fractional}
Kelly J~F and Meerschaert M~M 2019 The fractional advection-dispersion equation
  for contaminant transport {\em {H}andbook of {F}ractional {C}alculus with
  {A}pplications. {V}ol. 5\/} ed Tarasov V~E (De Gruyter, Berlin) pp 129--149

\bibitem{Yong2016Backward}
Zhang Y, Meerschaert M~M and Neupauer R~M 2016 Backward fractional advection
  dispersion model for contaminant source prediction {\em Water Resour. Res.\/}
  {\bf 52} 2462--2473
  \urlprefix\url{https://agupubs.onlinelibrary.wiley.com/doi/abs/10.1002/2015WR018515}

\bibitem{Deng2020Modeling}
Deng W, Hou R, Wang W and Xu P 2020 {\em Modeling Anomalous Diffusion: From
  Statistics to Mathematics\/} (Singapore: World Scientific) ISBN 9789811212994
  \urlprefix\url{https://books.google.com.hk/books?id=vUIQywEACAAJ}

\bibitem{Montroll1965Random}
Montroll E~W and Weiss G~H 1965 Random walks on lattices. {II} {\em J. Math.
  Phys.\/} {\bf 6} 167--181
  \urlprefix\url{http://scitation.aip.org/content/aip/journal/jmp/6/2/10.1063/1.1704269}

\bibitem{Kutner2017continuous}
Kutner R and Masoliver J 2017 The continuous time random walk, still trendy:
  fifty-year history, state of art and outlook {\em Eur. Phys. J. B\/} {\bf 90}
  50 ISSN 1434-6036 \urlprefix\url{https://doi.org/10.1140/epjb/e2016-70578-3}

\bibitem{Vot2020Continuous}
Vot F~L, Abad E, Metzler R and Yuste S~B 2020 Continuous time random walk in a
  velocity field: role of domain growth, {G}alilei-invariant
  advection-diffusion, and kinetics of particle mixing {\em New J. Phys.\/}
  {\bf 22} 073048 \urlprefix\url{https://dx.doi.org/10.1088/1367-2630/ab9ae2}

\bibitem{Berkowitz2006Modeling}
Berkowitz B, Cortis A, Dentz M and Scher H 2006 Modeling non-{F}ickian
  transport in geological formations as a continuous time random walk {\em Rev.
  Geophys.\/} {\bf 44} RG2003 (\textit{Preprint}
  \eprint{https://agupubs.onlinelibrary.wiley.com/doi/pdf/10.1029/2005RG000178})
  \urlprefix\url{https://agupubs.onlinelibrary.wiley.com/doi/abs/10.1029/2005RG000178}

\bibitem{Liu2020Correlated}
Liu J, Bao J~D and Chen X 2020 Correlated continuous-time random walk in a
  velocity field: Anomalous bifractional crossover {\em Phys. Rev. E\/} {\bf
  102}(6) 062122
  \urlprefix\url{https://link.aps.org/doi/10.1103/PhysRevE.102.062122}

\bibitem{Cherstvy2019Non}
Cherstvy A~G, Thapa S, Wagner C~E and Metzler R 2019 Non-gaussian{,}
  non-ergodic{,} and non-{F}ickian diffusion of tracers in mucin hydrogels {\em
  Soft Matter\/} {\bf 15}(12) 2526--2551
  \urlprefix\url{http://dx.doi.org/10.1039/C8SM02096E}

\bibitem{Akimoto2018Ergodicity}
Akimoto T, Cherstvy A~G and Metzler R 2018 Ergodicity, rejuvenation,
  enhancement, and slow relaxation of diffusion in biased continuous-time
  random walks {\em Phys. Rev. E\/} {\bf 98}(2) 022105
  \urlprefix\url{https://link.aps.org/doi/10.1103/PhysRevE.98.022105}

\bibitem{Wang2020Fractional}
Wang W and Barkai E 2020 Fractional advection-diffusion-asymmetry equation {\em
  Phys. Rev. Lett.\/} {\bf 125}(24) 240606
  \urlprefix\url{https://link.aps.org/doi/10.1103/PhysRevLett.125.240606}

\bibitem{Wang2019Transport}
Wang W, Vezzani A, Burioni R and Barkai E 2019 Transport in disordered systems:
  The single big jump approach {\em Phys. Rev. Res.\/} {\bf 1}(3) 033172
  \urlprefix\url{https://link.aps.org/doi/10.1103/PhysRevResearch.1.033172}

\bibitem{Wang2009Anomalous}
Wang B, Anthony S~M, Bae S~C and Granick S 2009 Anomalous yet {B}rownian {\em
  PNAS\/} {\bf 106} 15160--15164 ISSN 0027-8424
  \urlprefix\url{https://www.pnas.org/content/106/36/15160}

\bibitem{Chubynsky2014Diffusing}
Chubynsky M~V and Slater G~W 2014 Diffusing diffusivity: A model for anomalous,
  yet {B}rownian, diffusion {\em Phys. Rev. Lett.\/} {\bf 113}(9) 098302
  \urlprefix\url{https://link.aps.org/doi/10.1103/PhysRevLett.113.098302}

\bibitem{Pinaki2007Universal}
Chaudhuri P, Berthier L and Kob W 2007 Universal nature of particle
  displacements close to glass and jamming transitions {\em Phys. Rev. Lett.\/}
  {\bf 99}(6) 060604
  \urlprefix\url{https://link.aps.org/doi/10.1103/PhysRevLett.99.060604}

\bibitem{Barkai2020Packets}
Barkai E and Burov S 2020 Packets of diffusing particles exhibit universal
  exponential tails {\em Phys. Rev. Lett.\/} {\bf 124}(6) 060603
  \urlprefix\url{https://link.aps.org/doi/10.1103/PhysRevLett.124.060603}

\bibitem{Adrian2021Large}
Pacheco-Pozo A and Sokolov I~M 2021 Large deviations in continuous-time random
  walks {\em Phys. Rev. E\/} {\bf 103}(4) 042116
  \urlprefix\url{https://link.aps.org/doi/10.1103/PhysRevE.103.042116}

\bibitem{Defaveri2024Stretched}
Defaveri L, Barkai E and Kessler D~A 2024 Stretched-exponential relaxation in
  weakly confined brownian systems through large deviation theory {\em Phys.
  Rev. E\/} {\bf 109}(2) L022102
  \urlprefix\url{https://link.aps.org/doi/10.1103/PhysRevE.109.L022102}

\bibitem{defaveri2025diffusion}
Defaveri L and Barkai E 2025 Diffusion in quenched random environments:
  Reviving laplace's first law of errors (\textit{Preprint}
  \eprint{2501.05585}) \urlprefix\url{https://arxiv.org/abs/2501.05585}

\bibitem{Hidalgo2020Hitchhiker}
Hidalgo-Soria M and Barkai E 2019 The {H}itchhiker model for {L}aplace
  diffusion processes in the cell environment arXiv:1909.07189

\bibitem{hamdi2024laplace}
Hamdi O, Burov S and Barkai E 2024 Laplace’s first law of errors applied to
  diffusive motion {\em Eur. Phys. J. B\/} {\bf 97} 1--13

\bibitem{Burov2022Exponential}
Burov S, Wang W and Barkai E 2022 Exponential tails and asymmetry relations for
  the spread of biased random walks arXiv:2209.03410

\bibitem{Wang2026Far}
Wang W, Zhang K and Cheng Y 2026 Far tails of the biased continuous-time
  random-walk model in the short-time limit {\em Phys. Rev. E\/} {\bf 114}(2)
  024102 \urlprefix\url{https://link.aps.org/doi/10.1103/pn9s-lgxx}

\bibitem{Burioni2013Rare}
Burioni R, Gradenigo G, Sarracino A, Vezzani A and Vulpiani A 2013 Rare events
  and scaling properties in field-induced anomalous dynamics {\em J. Stat.
  Mech. Theory Exp.\/} {\bf 2013} P09022
  \urlprefix\url{http://stacks.iop.org/1742-5468/2013/i=09/a=P09022}

\bibitem{Godreche2001Statistics}
Godr{\`e}che C and Luck J~M 2001 Statistics of the occupation time of renewal
  processes {\em J. Stat. Phys.\/} {\bf 104} 489--524 ISSN 1572-9613
  \urlprefix\url{http://dx.doi.org/10.1023/A:1010364003250}

\bibitem{Wang2018Renewal}
Wang W, Schulz J~H~P, Deng W~H and Barkai E 2018 Renewal theory with fat-tailed
  distributed sojourn times: Typical versus rare {\em Phys. Rev. E\/} {\bf
  98}(4) 042139
  \urlprefix\url{https://link.aps.org/doi/10.1103/PhysRevE.98.042139}

\bibitem{Klafter2011First}
Klafter J and Sokolov I~M 2011 {\em {First Steps in Random Walks: From Tools to
  Applications}\/} (Oxford University Press, Oxford) ISBN 9780199234868
  \urlprefix\url{https://doi.org/10.1093/acprof:oso/9780199234868.001.0001}

\bibitem{Wang2026Statistics}
Wang W and Burov S 2026 Statistics of a large number of renewals in equilibrium
  and ordinary renewal processes at the short time limit {\em Chaos: An
  Interdisciplinary Journal of Nonlinear Science\/} {\bf 36} 043133 ISSN
  1054-1500 \urlprefix\url{https://doi.org/10.1063/5.0324747}

\bibitem{CorlessOn1996}
Corless R~M, Gonnet G~H, Hare D~E~G, Jeffrey D~J and Knuth D~E 1996 On the
  {L}ambert {$W$} function {\em Adv. Comput. Math.\/} {\bf 5} 329--359 ISSN
  1019-7168 \urlprefix\url{https://doi.org/10.1007/BF02124750}

\bibitem{Zarfaty2016Statistics}
Zarfaty L and Meerson B 2016 Statistics of large currents in the
  kipnis{\textendash}marchioro{\textendash}presutti model in a ring geometry
  {\em J. Stat. Mech: Theory Exp.\/} {\bf 2016} 033304
  \urlprefix\url{https://doi.org/10.1088/1742-5468/2016/03/033304}

\bibitem{Zarfaty2018Dispersion}
Zarfaty L, Peletskyi A, Fouxon I, Denisov S and Barkai E 2018 Dispersion of
  particles in an infinite-horizon lorentz gas {\em Phys. Rev. E\/} {\bf 98}(1)
  010101 \urlprefix\url{https://link.aps.org/doi/10.1103/PhysRevE.98.010101}

\bibitem{Touchette2009large}
Touchette H 2009 The large deviation approach to statistical mechanics {\em
  Phys. Rep.\/} {\bf 478} 1--69 ISSN 0370-1573
  \urlprefix\url{https://doi.org/10.1016/j.physrep.2009.05.002}

\bibitem{Wang2020Large}
Wang W, Barkai E and Burov S 2020 Large deviations for continuous time random
  walks {\em Entropy\/} {\bf 22} 697

\bibitem{Schneider1986Stable}
Schneider W~R 1986 {\em Stable distributions: {F}ox functions representation
  and generalization\/} ({\em Lecture Notes in Phys.\/} vol 262) (Springer,
  Berlin) pp 497--511 \urlprefix\url{https://doi.org/10.1007/3540171665_92}

\bibitem{Zolotarev1986One}
Zolotarev V~M 1986 {\em One-dimensional stable distributions\/} vol~65
  (American Mathematical Soc., Providence R.I.)

\bibitem{Schneider1989Fractional}
Schneider W~R and Wyss W 1989 Fractional diffusion and wave equations {\em J.
  Math. Phys.\/} {\bf 30} 134--144 ISSN 0022-2488
  \urlprefix\url{https://doi.org/10.1063/1.528578}

\bibitem{Barkai2003Aging}
Barkai E and Cheng Y~C 2003 Aging continuous time random walks {\em J. Chem.
  Phys.\/} {\bf 118} 6167--6178
  \urlprefix\url{http://aip.scitation.org/doi/abs/10.1063/1.1559676}

\bibitem{Schulz2014Aging}
Schulz J~H~P, Barkai E and Metzler R 2014 Aging renewal theory and application
  to random walks {\em Phys. Rev. X\/} {\bf 4}(1) 011028
  \urlprefix\url{https://link.aps.org/doi/10.1103/PhysRevX.4.011028}

\bibitem{bender1999advanced}
Bender C~M and Orszag S~A 1999 {\em Advanced mathematical methods for
  scientists and engineers: Asymptotic methods and perturbation theory\/} vol~1
  (Springer)

\end{thebibliography}

\end{document}